\documentclass[aps,floatfix,superscriptaddress,tightenlines,amsmath,amssymb,nofootinbib,10pt,longbibliography,twocolumn,prx]{revtex4-2}
\usepackage{lineno}
\usepackage[utf8]{inputenc}
\usepackage{graphicx}
\usepackage{amsfonts}
\usepackage{csquotes}
\usepackage{multirow} 
\usepackage{amsmath}
\usepackage{amsthm}
\usepackage{hyperref}
\usepackage{braket}
\usepackage{xcolor}
\usepackage{bbm}
\usepackage{comment}
\usepackage{enumitem}
\usepackage{bm} 
\usepackage{nicematrix}

\newtheorem{definition}{Definition}

\usepackage{wrapfig}
\usepackage{array}
\newcolumntype{L}[1]{>{\raggedright\let\newline\\\arraybackslash\hspace{0pt}}m{#1}}
\newcolumntype{C}[1]{>{\centering\let\newline\\\arraybackslash\hspace{0pt}}m{#1}}
\newcolumntype{R}[1]{>{\raggedleft\let\newline\\\arraybackslash\hspace{0pt}}m{#1}}

\begin{document}

\title{\normalfont\LARGE{
Quantum-private distributed sensing}}

\author{Joseph Ho}
\thanks{Corresponding email: joseph.ho@hw.ac.uk}
\affiliation{Institute of Photonics and Quantum Sciences, School of Engineering and Physical Sciences, Heriot-Watt University, Edinburgh EH14 4AS, United Kingdom}
\author{\hspace{-4pt}$^{,\dagger}$ Jonathan W. Webb}
\thanks{These authors contributed equally}
\affiliation{Institute of Photonics and Quantum Sciences, School of Engineering and Physical Sciences, Heriot-Watt University, Edinburgh EH14 4AS, United Kingdom}
\author{Russell M. J. Brooks}
\affiliation{Institute of Photonics and Quantum Sciences, School of Engineering and Physical Sciences, Heriot-Watt University, Edinburgh EH14 4AS, United Kingdom}
\author{Federico Grasselli}
\affiliation{Institut de Physique Th\'eorique, Universit\'e Paris-Saclay, CEA, CNRS, 91191 Gif-sur-Yvette, France}
\affiliation{Now at Leonardo Innovation Labs -- Quantum Technologies, Via Tiburtina km 12,400, 00131 Rome, Italy}
\author{Erik Gauger}
\affiliation{Institute of Photonics and Quantum Sciences, School of Engineering and Physical Sciences, Heriot-Watt University, Edinburgh EH14 4AS, United Kingdom}
\author{Alessandro Fedrizzi}
\affiliation{Institute of Photonics and Quantum Sciences, School of Engineering and Physical Sciences, Heriot-Watt University, Edinburgh EH14 4AS, United Kingdom}

\begin{abstract}
Quantum networks can enhance both security and privacy conditions for multi-user communication, delegated computation, and distributed sensing tasks.
An example quantum protocol is private parameter estimation (PPE) where only the aggregate information is accessible while individual sensor data remain confidential.
Specifically, the protocol enables the estimation of a global function of remote sensor parameters without revealing local parameters to any entity.
We implement the PPE protocol by distributing a three-photon Greenberger-Horne-Zeilinger (GHZ) state, among three sensors, which is verified using stabilizer measurements to establish privacy and precision bounds for the sensing task. We demonstrate Heisenberg-limited precision scaling of the global parameter while suppressing the metrological information of the local parameters by up to three orders of magnitude. This work, which integrates privacy in distributed quantum sensing, marks a crucial step towards developing advanced quantum-secure-and-private protocols in complex quantum networks.
\end{abstract}

\maketitle

\section{Introduction}
 \label{sec:intro}

Quantum networks allow connected nodes to perform tasks such as multi-user quantum cryptography~\cite{Gisin2002,Pirandola2020}, distributed quantum computation~\cite{barz2012demonstration,fitzsimons2017private}, and distributed quantum sensing~\cite{gottesman2012telescope,Komar2014clocks,baumgratz2016quantum,Liu2021,nichol2022clocks}.
The added connectivity and sharing of network resources may require privacy-enabled protocols that allow remote nodes to contribute to a task without learning privileged information.
A well-known example of a private computing protocol is blind quantum computing, which allows clients to access remote quantum processors without revealing the computational algorithm to the servers~\cite{Broadbent2009,barz2012demonstration,fitzsimons2017private}.
Another established application is anonymous quantum communication, in which users share cryptographic keys within a quantum network without revealing the identities of keyholders to an eavesdropper or other users within the network~\cite{hahn2020anonymous,thalacker2021anonymous,huang2022experimental,Grasselli2022,de2023anonymous,Webb24}.
Similar ideas from quantum cryptography can be incorporated into remote sensing tasks operating on quantum networks to provide security and privacy features~\cite{eldredge2018optimal,Huang2019,Shettell2022cryptographicMetrology,Kasai2023}.

\begin{figure}[t!]
    \centering
    \includegraphics[width=\linewidth]{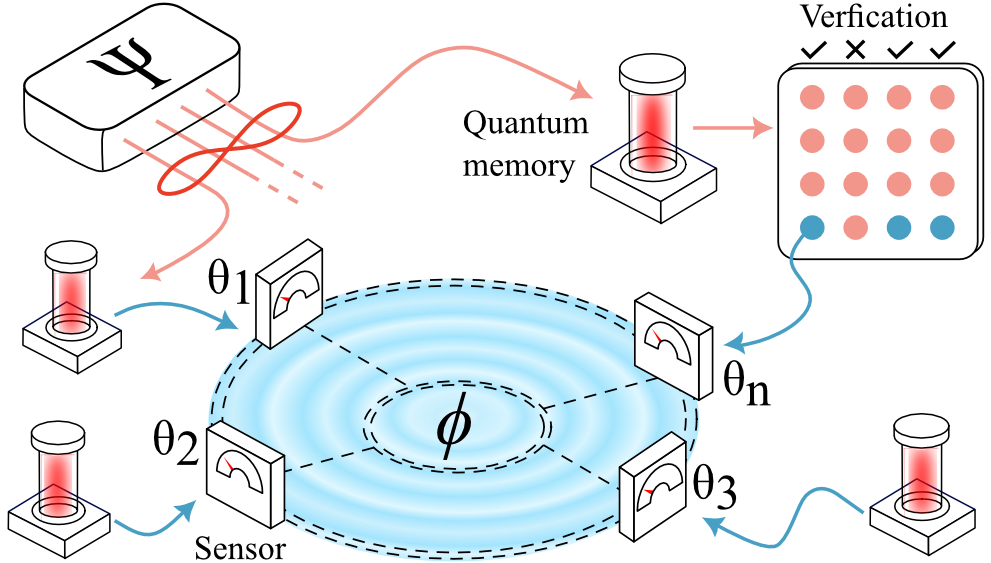}
    \caption{Illustration of private parameter estimation. 
    Sensors in a network monitor a global function of parameters while local values $\{\theta_i\}$ remain secret.
    An \textit{honest verifier} establishes private pairwise channels with each sensor (not shown).
    An \textit{untrusted server} sends $N_{\textrm{t}}$ copies of the resource state to all sensors, who store them locally in a quantum memory.
    They perform the verification protocol to ensure they share a state close to a GHZ state.
    If verification is successful, each sensor encodes their local parameter on one copy of the shared state to perform parameter estimation.
    }
    \label{fig:globe}
\end{figure}

Within this broader context, distributed quantum sensing has emerged as a key application of quantum networks.
Here, multiple quantum sensors collectively probe a global property of an interrogated sample to achieve higher precision than a single probe~\cite{zhang2021distributedRev}.
This differs from multi-parameter estimation, where each sensor value is estimated and local strategies are optimal~\cite{knott2016multiparameter}.
Distributed sensing is suited for non-localised phenomena, such as magnetic fields~\cite{mamin2013nanoscale}, radio-frequency sensing~\cite{xia2020demonstration}, and drifts between remote clocks~\cite{Komar2014clocks}. 
In these applications, the goal is to estimate a global value $\phi$ that is a linear function of local sensor values $\phi = \boldsymbol{w \cdot \varphi}$, where $\boldsymbol{w}$ are the weights and $\boldsymbol{\varphi}$ is an array of parameters~\cite{ge2018distributed,zhuang2018distributed,proctor2018multiparameter}.
When estimating the weighted average, distributed sensing with entanglement can surpass the standard quantum limit and achieve Heisenberg-limited precision scaling~\cite{eldredge2018optimal,qian2019heisenberg,Liu2021}.
Additionally, entanglement can be used to guarantee security, ensuring that only authorized users have access to the measured parameters, and privacy -- delegating the sensing task to an untrusted sensor to perform measurements without learning the parameters~\cite{Huang2019,Shettell2022cryptographicMetrology}.
Recently, a two-user scheme has been proposed~\cite{Takeuchi2019} and demonstrated~\cite{Yin2020} by sharing a Bell pair between two parties, one user performs the sensing task while the other learns the estimated parameter.

In this work we consider a multi-user quantum-private distributed sensing task called \emph{private parameter estimation} (PPE) for which security bounds were recently established in Ref~\cite{Shettell2022arxiv}.
As shown in Fig.\ref{fig:globe}, this involves $n$ remote sensors that contribute to estimating a global phase $\bar{\phi}$, which is the weighted average $w_i=1/n$ of local sensor phases $\theta_i$, with $i \in \{1,\dots, , n\}$.
The goal is to prevent information about $\theta_i$ from being learned by any participant or eavesdropper monitoring the communication, while still measuring $\phi$ with Heisenberg-limited precision.
This notion of privacy differs from the two-user delegated sensing tasks~\cite{Takeuchi2019,Yin2020} since those protocols allow one user to fully learn the other's sensor value, while in PPE all local sensor information is hidden and only the aggregated value is known.
Analogous to quantum conference key agreement~\cite{Proietti2021,pickston2023conference,Webb24,hahn2020anonymous,Murta2020}, the correlations of an $n$-partite GHZ state, $\ket{GHZ} \doteq (\ket{0}^{\otimes{n}} + \ket{1}^{\otimes{n}})/\sqrt{2}$ where $\ket{0}$ and $\ket{1}$ are computational basis states, are exploited to perform private distributed sensing.

\section{Methods}
\subsection*{Private sensing protocol}

We briefly outline the PPE protocol~\cite{Shettell2022arxiv} illustrated in Fig.~\ref{fig:globe}.
An honest verifier initiates the protocol by establishing private pairwise channels with $n$ sensors.
A quantum server, which can be untrusted, prepares and distributes $N_t$ copies of the $n$-partite GHZ state to the sensors who store all copies in quantum memories.
The verifier randomly picks half of the copies ($N_t/2$) and instructs the group to measure their respective copies with respect to the stabilizers, $K_i$, which correspond to local $X$ or $Y$ basis measurements~\cite{toth2005entanglement} (more details in the Supplementary Materials~\ref{sec:protocols}).
Each sensor performs their measurement then communicates the outcome to the verifier directly using pairwise private channels.
The verifier computes a failure rate, $f$, and if $f \leq 1/(2n^2)$, where $1/(2n^2)$ is the maximum allowed failure rate~\cite{Shettell2022arxiv}, then one remaining untested copy is randomly chosen as the target copy, $\tau$, for the parameter estimation task.
This verification scheme is related to the scheme in Ref.~\cite{unnikrishnan2022verification} which can verify any graph state but requires $2n$ stabilizer measurements for an $n$-qubit graph.
By comparison, this protocol uses $n+1$ measurements to verify a $n$-partite GHZ state~\cite{Shettell2022arxiv}.
This verification protocol guarantees a lower bound on the state fidelity $F(\tau,\ket{GHZ})$ of the target copy with a GHZ state at a nominal probability,
\begin{equation}
    \mathbb{P}\bigg(F(\tau,\ket{GHZ}) \geq 1-\frac{2\sqrt{c}}{n}-2nf\Bigg)\geq 1-n^{1-\frac{2mc}{3}}.
    \label{eq:verification-prob}
\end{equation}
$\mathbb{P}(\textbf{A})$ is the probability that $\textbf{A}$ is true, given the failure rate $f$, number of sensors $n$, and two positive variables $c$ and $m$ which relate to the statistical certainty based on the total number of copies, $N_t = \lceil 2mn^5\text{log}(n) \rceil$, used in the protocol (more details in the Supplementary Materials~\ref{sec:protocols}).
Quantum memories prevent an adversary swapping out non-test copies, after distribution, with non-private states such as $|+\rangle^{\otimes{n}}$ where $|+\rangle \doteq (|0\rangle+|1\rangle)/\sqrt{2}$ which would allow the adversary to learn the local phases from the outcomes.
We assume the future availability of such quantum memories and prepare multi-partite GHZ states to investigate the implementation of the verification scheme presented in Ref.~\cite{Shettell2022arxiv}.

If the verification protocol is successful, the verifier announces one target copy, $\tau$, to be used to encode the local phases and perform parameter estimation.
Using the measured $f$, one constructs a bound on the privacy using the continuity of Fisher Information~\cite{continuityFisherInfo} which provides an upper bound on $\varepsilon_p$ given as\cite{Shettell2022arxiv},\footnote{We corrected the equation from Ref.\cite{Shettell2022arxiv}, $\varepsilon_p \leq  \frac{24}{\tilde{n}}\sqrt{\frac{2\sqrt{c}}{n} - 2nf}$},
\begin{equation}
    0 \leq \varepsilon_p \leq  \frac{24}{n^2}\sqrt{\frac{2\sqrt{c}}{n} + 2nf},
    \label{eq:privacy}
\end{equation}
where in the ideal case, $\varepsilon_p=0$, implies no information can be learned about the local sensor parameters $\theta_i$.

The remaining $N_t/2-1$ untested copies are discarded\footnote{Note that it might be possible to use some of these discarded copies in the protocol, however the security implications of this are not yet clear.},
which is needed to satisfy the security demands for the verification scheme~\cite{Shettell2022arxiv}.
The private sensing task proceeds with each sensor applying local unitary encoding operators, $U(\theta_i)=\exp(-i \theta_i Z/2)$, where $Z$ is the Pauli-Z matrix, on their qubit.
Finally, each sensor measures in the $X$ basis and announces their outcome.
The parity of the outcomes is used to estimate $\phi$ by repeating the protocol $\nu$ times.
We use $\phi$ to estimate the average phase of the local sensors, which obtains Heisenberg-limited scaling $\text{var}[\phi] \propto (N \nu)^{-1}$.

\subsection*{Experimental setup}
\label{sec:prepGHZstate}
\begin{figure}
    \centering
    \includegraphics[width=\linewidth]{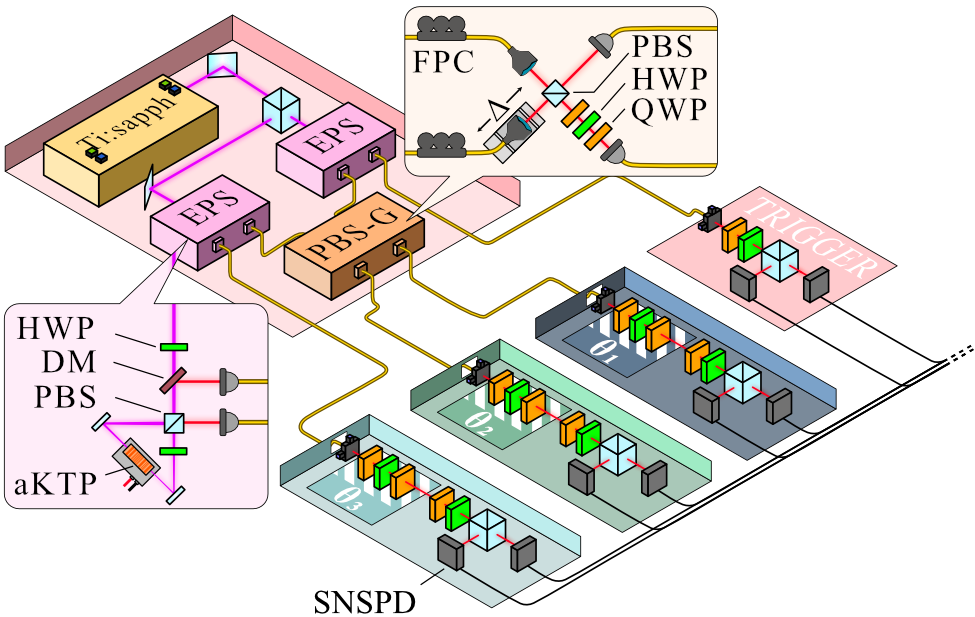}
    \caption{Experimental setup preparing a three-qubit GHZ state and implementing PPE.
    A Ti:sapph laser pumps two entangled photon-pair sources (EPS). One photon from each source is sent to a polarising beamsplitter gate (PBS-G). Each sensor has a local phase encoding stage ($\theta_i$). See Methods for detail. HWP: half-wave plate, QWP: quarter-wave plate, PBS: polarising beamsplitter, DM: dichroic mirror, FPC: fibre polarisation controller, $\Delta$: delay stage, aKTP: aperiodically-poled KTP crystal, SNSPD: superconducting nanowire single photon detector.
    }
    \label{fig:experiment}
\end{figure}

We experimentally prepare a three-qubit GHZ state to investigate the PPE protocol for $n=3$ sensors as shown in Fig.~\ref{fig:experiment}.
Both EPS produce spectrally-pure entangled photon pairs using a tailored aperiodically-poled KTP crystal~\cite{pickston2021optimised} in a Sagnac configuration~\cite{fedrizzi2007}.
Using the PBS-G we generate the three-photon GHZ state, conditional on measuring one photon in each mode.
All four photons are coupled into single-mode fibre and sent to the sensor stages, with the forth photon measured as the trigger.

The sensor nodes consist of a QWP-HWP-QWP, which allows for encoding local phases $\theta_i$, and this is set up before a polarisation analyser consisting of a QWP, HWP then PBS.
We use SNSPDs to detect single photons in both outputs of each PBS.
This allows for tomography measurements and the protocol measurements respectively.
For more details on the setup and its characterisation see Supplementary Material~\ref{sec:experiment}.

\section{Results}
\subsection*{Evaluating Privacy}

\begin{figure}
    \centering
    \includegraphics[width=\linewidth]{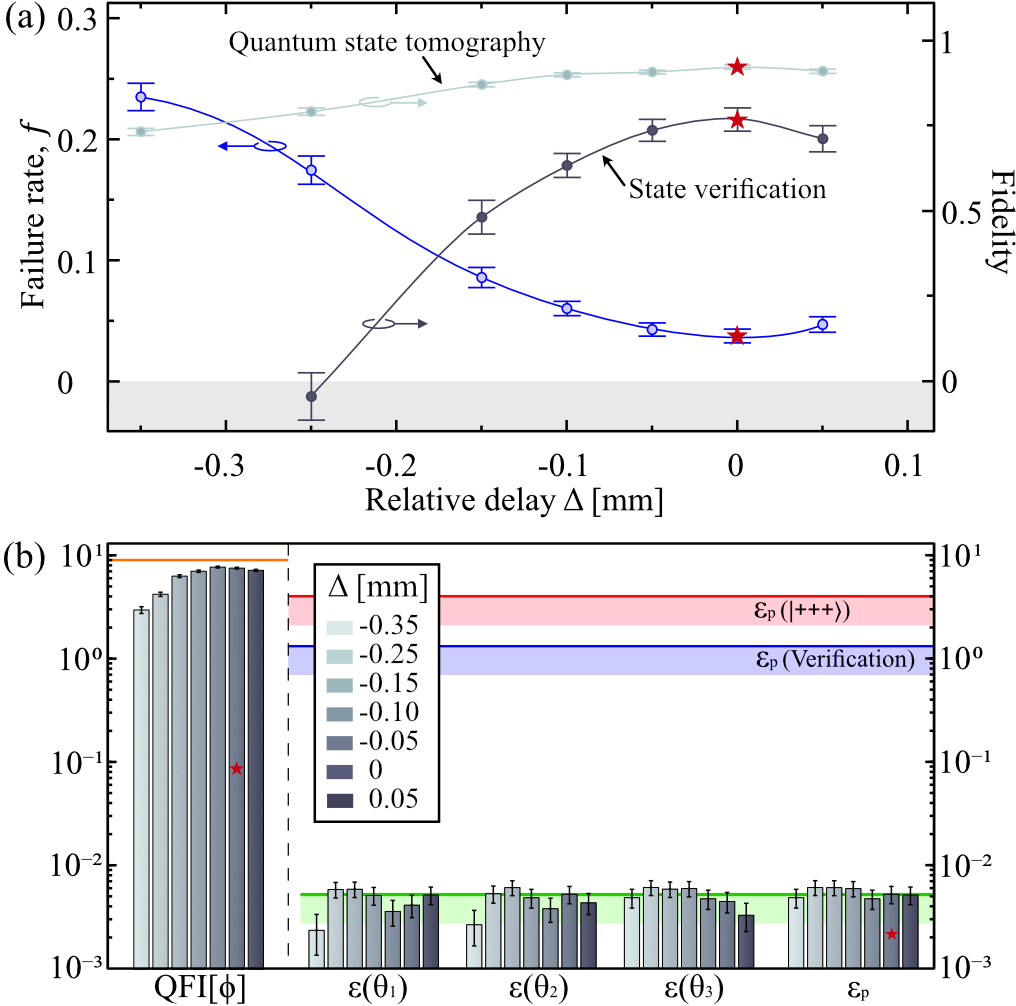}
    \caption{Verification protocol results.
    (a) Using stabilizer measurements the failure rate $f$ is measured (dark blue dots) for each delay, $\Delta$ then we evaluate the lower bound fidelity (dark grey dots).
    We perform quantum state tomography, reconstruct the density matrices, then estimate the GHZ state fidelity (light blue dots).
    Solid lines are a visual guide for datasets and not fits.
    (b) Using the density matrices we directly calculate QFI for estimating $\phi$ for each $\Delta$ which has a maximum QFI~$=9$ (orange line).
    We directly calculate privacy parameters $\varepsilon(\theta_i)$ and protocol privacy $\varepsilon_p = \max_{i}{[\varepsilon (\theta_i)]}$ from the density matrices.
    From the verification procedure we obtain an upper-bound on $\varepsilon_p$ (blue line) for our optimal state.
    We evaluate the upper-bound for $|+++\rangle$ (red line) using the verification process.
    For comparison, the green line is the directly evaluated $\varepsilon_p$ from the density matrices.
    All error bars derived from Monte Carlo sampling with 200 samples, assuming Poissonian statistics.
    Red star denotes the optimal setting, $\Delta=0$.
    }
    \label{fig:results}
\end{figure}

We investigate the verification protocol by considering three main properties; the robustness to noisy GHZ states, comparing the privacy bounds with a direct evaluation of $\varepsilon_p$, and the impact of finite copy statistics.
To test the robustness of the verification protocol we prepare GHZ states with noise added by setting the $\Delta$ away from 0 and compute $f$ in each case, see Fig~\ref{fig:results}(a).
For each state we perform quantum state tomography (QST) to directly obtain fidelity values.
We also show the lower-bound fidelity using the stabilizers in Fig.~\ref{fig:results}(a).
In all cases, the lower bound is much smaller than the measured fidelity value which shows the protocol obeys correctness as it never overestimates the fidelity.

To demonstrate privacy of $\theta_i$, we use definitions of privacy introduced in Ref.~\cite{Shettell2022arxiv} based on the quantum Fisher information (QFI), a standard tool for benchmarking quantum metrology schemes.
This uses density matrices $\rho$ of the state we obtained. 
In the optimal configuration ($\Delta=0$) the three-qubit GHZ state fidelity was $0.923\pm0.005$ with a state purity of $0.865\pm0.009$.
Using this reconstructed density matrix, we directly compute QFI$=7.7\pm0.2$ for measuring $\phi$, while for an ideal GHZ state we expect QFI~$=9$, see Fig~\ref{fig:results}(b).
We evaluate the protocol privacy as $\varepsilon_p=0.005\pm0.002$ by calculating the \emph{reducible} QFI for each $\theta_i$, when encoded by the $i$-th sensor and when the other $j \neq i$ sensors encode functions of $\theta_i$ such that the resulting QFI is minimised, see Supplementary Materials for details.
When implementing the PPE protocol, we measure stabilizers $K_i$, which are a subset of the recorded QST measurements.
The average failure rate was $f=0.039\pm0.005$, which is below the abort threshold value, i.e., $f\leq1/(2n^2)=0.05\dot{5}$~\cite{Shettell2022arxiv} for $n=3$.
We then use $f$ to establish a lower-bound fidelity of $0.769\pm0.006$, via $F(\tau,GHZ)\geq 1-2nf$ when assuming infinite resource-state copies ($c\rightarrow0$), which is smaller than the fidelity obtained via QST.
Finally we establish an upper-bounded privacy parameter using Eq~\ref{eq:privacy}, $\varepsilon_p \leq 1.3 \pm 0.2$, again assuming infinite copies ($c\rightarrow 0$) are available. 
We find the upper bound is two orders of magnitude greater than the direct evaluation of $\varepsilon_p$.
We also calculate the expected $\varepsilon_p$ if a non-private state, e.g., $|+++\rangle$ was used, by evaluating the expected failure rate then computing $\varepsilon_p=4$.

Our main results consider the PPE task in the asymptotic regime assuming infinite copies, however in practice only finite copies will be possible.
The impact of a finite $N_t$ value is that $c$ does not become zero and acts as a statistical correction term to the bounds on fidelity and privacy; specifically $2\sqrt{c}/n$.
As an example, if we require that the definitions on the bounds hold with a failure rate below $10^{-6}$, this requires $2\times10^7$ total copies to achieve the correction term below $0.02$, see Supplementary Materials~\ref{sec:protocols}.

\subsection*{Evaluating phase estimation}
\begin{figure}
    \centering
    \includegraphics[width=\linewidth]{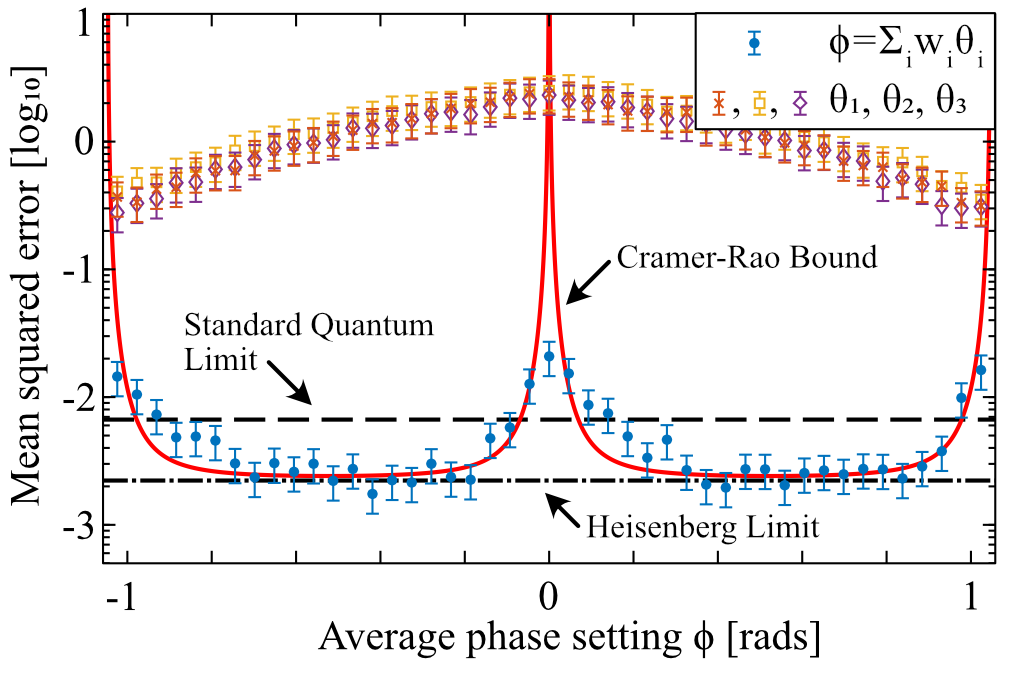}
    \caption{Private parameter estimation spanning the parameter space, $\phi \in [-1.05, 1.05]$ for a sample size $\nu=50$ using $n=3$ sensors.
    To collect statistics and evaluate the mean squared error, each estimation scheme is repeated 100 times.
    Solid circle data points represent evaluations of the mean value of the global phase $\phi$, while the other markers represent the evaluation of individual sensor phase values $\theta_i$.
    The red solid line is the theoretical Cramer-Rao bound obtained from estimating the Fisher information of the estimator with non-unit visibility.
    The dashed lines represent the theoretical lower-bounds based on the standard quantum limit precision scaling and Heisenberg limit scaling respectively.
    Error bars for all data points are evaluated using standard error, with three standard deviations shown.}
    \label{fig:parameterEst}
\end{figure}

We evaluated the performance of the private parameter estimation task over the range of phases $\phi\in[-1.05,1.05]$ (radians) as shown in see Fig.~\ref{fig:parameterEst}.
To estimate the global phase, $\phi$, all sensor measurement stages measure in $X$ then using the expectation value, we evaluate $\phi=\textrm{cos}^{-1}[ \langle X^{\otimes n}\rangle]$.
In the case of estimating the global parameter, specifically when evaluating the average local phase value $\phi\doteq \bar{\theta_i}= N^{-1}\Sigma_i^N \theta_i$, we find the mean square error has a phase dependent performance.
The mean squared error is evaluated as $\mathrm{MSE}(\hat{\phi}) =  \sum^n_{i=1} (\phi - \hat{\phi}_i)^2 /n$.
This is due to the limited visibility when measuring the $XXX$ of the GHZ state, however we observe regions of $\phi$ where measurements can achieve errors below the standard quantum limit and approach the Heisenberg limit.
In contrast, direct estimation of local phase terms consistently yields estimators with two orders of magnitude larger mean squared error than estimators on the global phase.
Additionally, the phase sensitivity is unaffected by the phase setting, i.e., it is uniformly insensitive to the global phase value.
This demonstrates the ability to use the PPE protocol to perform phase sensing at the ultimate precision limit, for certain global evaluations of local phases, while ensuring the information gain on the local sensor values are limited.
We note that the privacy on the local sensor values are not perfect, as there are slight biases which indicate that for a sufficient number of samples it is possible to learn some information about the local values.

\section{Discussion}

We have demonstrated a quantum-private distributed sensing task involving three sensors that measure a global phase $\phi$ without revealing local sensor values $\theta_i$---the precision in estimating $\phi$ is at least two orders of magnitude better than any $\theta_i$.
We assumed the availability of pairwise private channels between each sensor and the honest verifier for securely reporting outcomes in the verification protocol, this is a common requirement in other privacy-enabled protocols~\cite{thalacker2021anonymous,hahn2020anonymous,Grasselli2022,Webb24,de2023anonymous}.
Our work highlights a number of outstanding challenges in the implemented PPE protocol.
First, we found the bounds on the fidelity of the GHZ state and privacy parameter of the task, when using the failure rate in the verification protocol, are rather loose.
Tighter security, for example by rethinking the privacy definition in a composable security framework~\cite{Colisson2024}, would make distributed network sensing more resource efficient, and there is no fundamental reason preventing these from being derived in the future.
Second, in its current form, the PPE protocol requires all copies of the shared states to be stored in memories before they can be used.
The finite bandwidth of multi-mode memories would strongly limit the number of copies that can be used, and as we have seen from the performance penalties of our finite round analysis, PPE would effectively become impractical.
One may alleviate the need for large memories by introducing a trusted random variable that chooses whether a distributed state in a sequence is used or not for verification, such that each sensor only requires a single-qubit quantum memory~\cite{Unnikrishnan2019}.
Alternatively, one may consider GHZ-state certification protocols which exploit fast, low-loss optical switches to randomly select a copy from an ensemble which can be used for parameter estimation, while the remaining copies are certified~\cite{govcanin2022sample,antesberger2024efficient,martins2024experimental}.
Another improvement required to scale to more sensors is to revisit the current $1/n^2$ scaling for the fidelity bounds required for validation.
Regarding the resource state, it was recently proven in Ref.~\cite{bugalho2025,hassani2025} that the GHZ state, and equivalents up to a local unitary, is the only private state for estimating certain linear functions of inputs which our task falls within.
While the PPE protocol we have demonstrated has caveats, we also identified pathways for improvements, which, once addressed, will unlock a wider range and more efficient distributed sensing applications for quantum networks.

\vspace{1em}
\noindent
\textbf{Acknowledgements}
\newline
We thank A. Pickston, L. Stroh, D. Markham and N. Shettell for helpful discussions.
This work was supported by the UK Engineering and Physical Sciences Research Council (Grant Nos. EP/T001011/1.).
F. Grasselli acknowledges funding by the European Union’s Horizon Europe research and innovation program under the project “Quantum Security Networks Partnership” (QSNP, Grant Agreement No. 101114043) and “Quantum Internet Alliance”  (QIA-Phase 1, Grant Agreement No. 101102140).
FG did not contribute to this work on behalf of Leonardo S.p.A.

\vspace{1em}
\noindent
\textbf{Author contributions}
\newline
JH and JWW are co-first authors of this work.
JWW and AF conceived the project. 
JH, JWW, RMJB performed the experiment which includes taking data and analysing the results.
FG and EG developed the theoretical tools used in the analysis.
All authors contributed to writing and revisions of the manuscript.

\vspace{1em}
\noindent
\textbf{Competing interests}
\newline
The authors declare no competing financial or non-financial interests.

%


\onecolumngrid
\clearpage
\newpage

\setcounter{equation}{0}
\setcounter{figure}{0}
\setcounter{section}{0}
\makeatletter

\setcounter{secnumdepth}{1}

\renewcommand{\thefigure}{SM\arabic{figure}}
\renewcommand{\theequation}{SM\arabic{equation}} 
\renewcommand{\thesection}{\arabic{section}}  

\nolinenumbers

\pagestyle{empty}
\begin{center}
    \textbf{\large \label{SM}Supplemental Materials}
\end{center}

\setcounter{section}{0}

\section{Private parameter estimation protocols}
\label{sec:protocols}

The protocol we implement was introduced in Ref.~\cite{Shettell2022arxiv} as two separate protocols called \textsc{Verification} and \textsc{Secure network sensing} which we reproduce here for completeness.
Here we will refer to the first protocol as \textsc{State Verification}, to clarify that it certifies the state rather than a property of the system such as entanglement.
The structure of the protocol is shown in Table~\ref{tab:VerificationProtocolOne}.

\begin{table}[h!]
\centering
\begin{NiceTabular}{>{\raggedright\arraybackslash}p{0.3cm} >{\raggedright\arraybackslash}p{0.9\linewidth}}
\hline
\multicolumn{1}{c}{\space }  & \multicolumn{1}{c}{\textbf{Protocol 1}: \textsc{State Verification}~\cite{unnikrishnan2022verification}} \\
\cline{1-2}\\
1: \space & The verifier initiates the protocol with $n$ sensors indexed by $j \in \{1,2,...,n\}$.\\
2: & The verifier requests an non-trusted quantum server to distribute $N_t$ copies of the GHZ state to the $n$ sensors, each storing the copies locally.\\
3: & For each stabilizer $K_i$, where $i \in \{1,2,...,(n+1)\}$, repeat the following:  \begin{itemize}
        \item[(a)] The verifier selects $N_t/(2n+1)$ copies independently and uniformly at random.
        \item[(b)] For each copy, the verifier instructs each sensor measure their qubit according to $K_i$.
        \item[(c)] Each sensor sends their outcome to the verifier who computes $N_{\text{pass},i}$ which is the number of copies resulting in a $+1$ eigenvalue for the joint measurement outcome of $K_i$.
\end{itemize} \\
4: &  The verifier randomly chooses one of the $N_t/2$ remaining unmeasured copies to be the target copy, $\tau$.\\
5: &  The average failure rate $f=1-2\Sigma_i(N_{\text{pass},i})/N_t $ is calculated.\\
& \textit{if} $f\leq \frac{1}{2n^2}$, the parties use the target copy $\rho$ for their task; \\
& \textit{else} the verification of the GHZ state failed, the target copy is discarded and the protocol aborts.\\
\cline{1-2}
\end{NiceTabular}
\caption{\textsc{State Verification} protocol assuming the verifier is honest. \\The set of $(n+1)$ stabilizers are given by $K_{i=\{1,2,...,n\}}=- X^{(1)}X^{(2)} \dots Y^{(i)}Y^{(i+1)}\dots X^{(n-1)}X^{(n)}$ 
and $K_{i=n+1} = X^{\otimes n}$~\cite{toth2005entanglement}.\\
\textit{Input}: The parties choose the total number of copies of the resource state to be distributed, $N_t$, and the acceptable failure $f$ provided it obeys $f \leq \frac{1}{2n^2}$ to determine the lower-bounded fidelity to be certified.\\
\textit{Output:} a target copy $\tau$ close to the GHZ state or abort.}
\label{tab:VerificationProtocolOne}
\end{table}

The \textsc{State Verification} protocol allows an honest verifier to determine whether a distributed state is close to the ideal $n$-qubit GHZ state which is defined by the lower-bounded fidelity.
The verifier does not have to be holding one of the quantum sensors, instead they can delegate the sensing task to a quantum network of sensors.
The verifier is required to have pairwise private channels with each of the $n$ nodes prior to initiating the protocol.

\begin{figure}[h!]
    \centering
    \includegraphics[width=\linewidth]{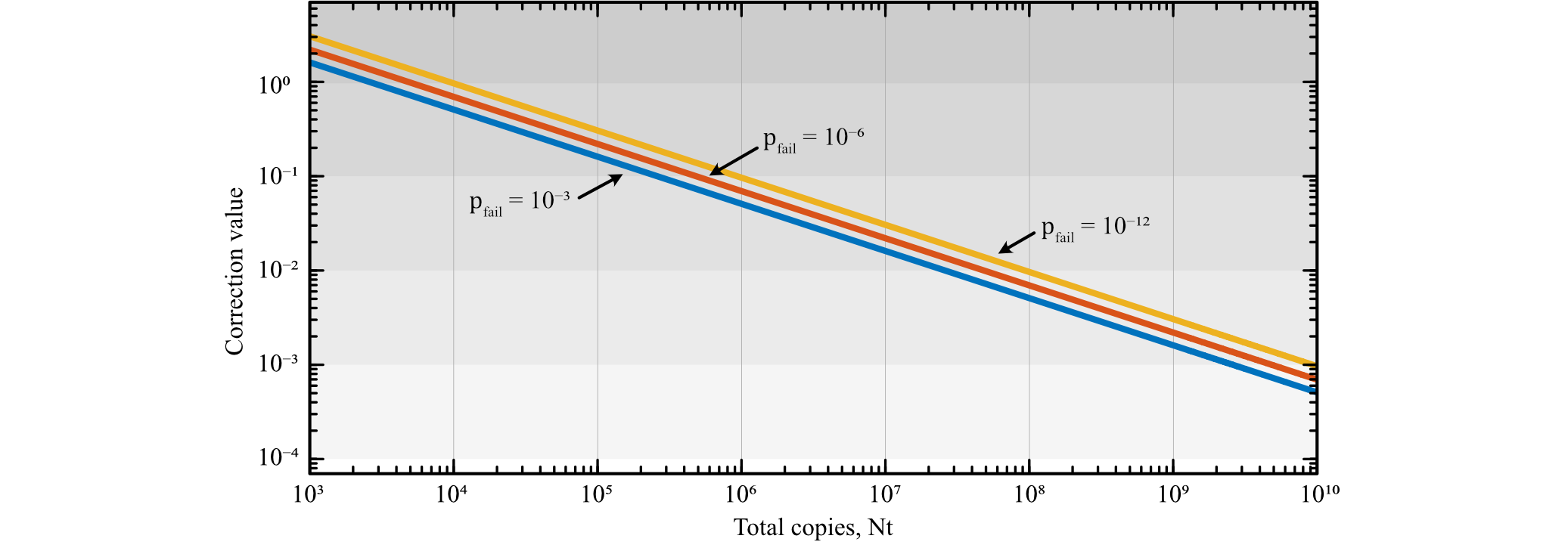}
    \caption{The correction term, $2\sqrt{c}/n$, is evaluated for a range of total copies $N_t$ used in the verification stage for $n=3$. 
    This is shown for three values of failure probabilities $p_{fail}$, which corresponds to the liklihood that the verification procedure fails to correctly lower bound the fidelity and upper bound $\varepsilon_p$.}
    \label{fig:correctionTerm}
\end{figure}

The first step is to decide on the total number of copies, $N_t$, to be used in the protocol as this will set the statistical confidence on the certified copy.
Notably, this is influenced directly by the two positive parameters $c$ and $m$ which themselves are related through $3/(2m) < c < (n-1)^2/4$.
Furthermore, the parameter $m$ is related to the total number of copies used in the protocol via $N_t = \lceil 2mn^5\text{log}(n) \rceil$.
To relate these parameters to the \textsc{State Verification} protocol, let us re-state the relevant equations,
\begin{equation}
    \mathbb{P}\bigg(F(\tau,\ket{GHZ}) \geq 1-\frac{2\sqrt{c}}{n}-2nf\Bigg)\geq 1-n^{1-\frac{2mc}{3}} \label{SM_fid}
\end{equation}
which establishes the lower-bound on the fidelity of the state.
We also have,
\begin{equation}
    \mathbb{P}\bigg(\varepsilon \geq \frac{24}{n^2}\sqrt{\frac{2\sqrt{c}}{n}+2nf}\Bigg)\geq 1-n^{1-\frac{2mc}{3}}
\end{equation}
which is related to the privacy definition in Ref~\cite{Shettell2022arxiv}. The $1/n^2$ scaling comes from the fact that we must account for the maximum QFI for an $n$-partite GHZ state, which bounds the private information via $F_{Q|\theta}(\rho_\theta) \leq \varepsilon_p n^2$. These two equations share similarities in their construction, they are both statements containing the probability $\mathbb{P}(\textbf{A})$ that the definition $\textbf{A}$ holds true is given by $1-n^{1-\frac{2mc}{3}}$.
Equivalently we can therefore infer that the probability that the definition $\textbf{A}$ fails is given by $p_{fail}=n^{1-\frac{2mc}{3}}$.
This can be a useful metric to enforce a certain level of confidence that the definition holds.
The probabilistic nature of these definitions are the result of only a subset of copies being measured.
Another observation to make is that $f$ is a critical parameter which is the measured paramter from the verification procedure, and sets how low the lower-bound on fidelity is, and high high the upper-bound on $\varepsilon_p$.
It stands that for smaller values of the measured failure rate, yields improved bounds in each case.
In addition to this, we identify a correction value of the form $\frac{2\sqrt{c}}{n}$ which plays a commensurate role in each parameter.
In general, the goal is to obtain the smallest value of $c$, to minimise this term, however as $c$ is bounded by $m$ which is set by the total number of copies, we obtain a direct relationship which is captured in Fig.~\ref{fig:correctionTerm}.

We consider a reasonable value for the correction term to be on the order of $10^{-2}$ or smaller, since the failure rate threshold scales as $\frac{1}{2n^2}$.
If we also require that the probability of the definition failing is $10^{-6}$, we find that at least approximately $2\times10^7$ copies would be needed.
It is clearly beneficial to choose a large value for $N_t$, however the constraints will likely be the size of the quantum memory.

\textsc{Secure network sensing} is a protocol which conditionally implements the private distributed-sensing task if \textsc{State Verification} is successful.
Depending on the chosen parameters $N_t$ and $f$ which set the lower-bound fidelity, and the upper-bound privacy condition (see Supplemental Materials~\ref{sec:privacy_an_integrity} for details), this can involve iterating \textsc{State Verification} until a target copy of the GHZ state is shared.
This is then repeated $\nu$ times in order to achieve a precision desired.
While the two protocols have been separately defined they are to be used together in order to carry out the PPE protocol.
The general structure of the protocol is given in Table~\ref{tab:protocol2}.

\begin{table}[h!]
\centering
\begin{NiceTabular}{>{\raggedright\arraybackslash}p{0.3cm} >{\raggedright\arraybackslash}p{0.9\linewidth}}
\hline
\multicolumn{1}{l}{\space}    & \multicolumn{1}{c}{\textbf{Protocol 2}: \textsc{Secure network sensing}~\cite{Shettell2022arxiv}} \\
\cline{1-2}\\
1:  & Repeat the following $\nu $ times.
\begin{itemize}
    \item[(a)] The sensors run \textsc{State Verification} until a GHZ state is shared.
    \item[(b)] Each sensor $j$ encodes their local parameter $\theta_j$ by applying $U_j = e^{-i\theta_j Z/2}$ on their qubit.
    \item[(c)] Each sensor measures their qubit in the $X$ basis then announces the measurement outcome.
\end{itemize}\\
2:  & The outcomes are used to compute the expectation value, $\langle X^{\otimes n} \rangle$, then the global phase is evaluated as $\hat{\phi} = \textrm{cos}^{-1}[\langle X^{\otimes n} \rangle]$. \\
\cline{1-2}
\end{NiceTabular}
\caption{\textsc{Secure network sensing} protocol \cite{Shettell2022arxiv}. \textit{Input:} The parties choose a level of precision for estimating $\phi$ which sets the number of rounds $\nu$. Each sensor encodes local phases  $\theta_1, \ldots, \theta_n$.\\
\textit{Output:} Estimation of $\phi=\sum_i \theta_i$, with $\varepsilon_i$-integrity and $\varepsilon_p$-privacy.}
\label{tab:protocol2}
\end{table}

\newpage
\section{Experimental setup}
\label{sec:experiment}
A mode-locked Ti:sapph laser operating at 80~MHz repetition rate, pulse duration of 1.3~ps, and central wavelength at 775~nm, optically pumps two EPS to generate entangled photon pairs.
Each EPS uses a 30~mm long aperiodically-poled KTP crystal that is quasi-phase-matched for Type-II parametric down-conversion and domain engineered to produce spectrally pure biphoton states~\cite{pickston2021optimised} without using narrowband filters while long-pass filters (not shown) are used to remove pump light.
We mount each aKTP crystal in a temperature-controlled oven to maximise the spectral overlap of the two down-converted photons, nominally centred at 1550~nm.
The aKTP crystal is embedded in a Sagnac configuration~\cite{fedrizzi2007}, consisting of a dual-wavelength PBS, dual-wavelength HWP, two silvered mirrors and a dichroic mirror which reflects 1550~nm photons and transmit 775~nm pump.
We use a HWP to prepare diagonally-polarised pump to create polarisation entanglement, e.g., $(\ket{hv}+\ket{vh})/{\sqrt{2}}$, where $\ket{h}\doteq\ket{0}$ and $\ket{v}\doteq\ket{1}$ are polarisation-encoded single photons.
Each photon is coupled in single-mode fibres using aspheric coupling lenses.
Each EPS achieved an average symmetric heralding efficiency of 50~\% with a source brightness of 1800~pairs/sec/mW.
In the experiments we pumped each source with 100~mW which ensures a source fidelity of 97.7~\% was obtained.

We use a polarising beamsplitter gate (PBS-G) to prepare the GHZ state in post-selection with a success probability of 50\%.
This consists of overlapping one photon from each source on a PBS, then conditional on measuring one photon at each output port of the PBS we obtain,
\begin{equation}
    \ket{GHZ} = \frac{\ket{hhhh} + e^{i\vartheta}\ket{vvvv}}{\sqrt{2}},
\end{equation}
where $\vartheta$ is an optical phase compensated by the QWP-HWP-QWP, where the each QWP is set to $+45^\circ$ from optic axis, and the HWP is rotated to compensate this phase.
To obtain the correct form of the GHZ state, the polarisation of the two input photons is set with fibre polarisation controllers (FPCs).

The four photons are measured using polarisation analysers which consist of a QWP, HWP and PBS, allowing arbitrary projective measurements on each photon.
While the PPE protocol only require $X$ and $Y$ measurements, the polarisation analysers allows for full QST to reconstruct the density matrix of the prepared state.

\begin{figure}[h!]
    \centering
    \includegraphics[width=0.95\linewidth]{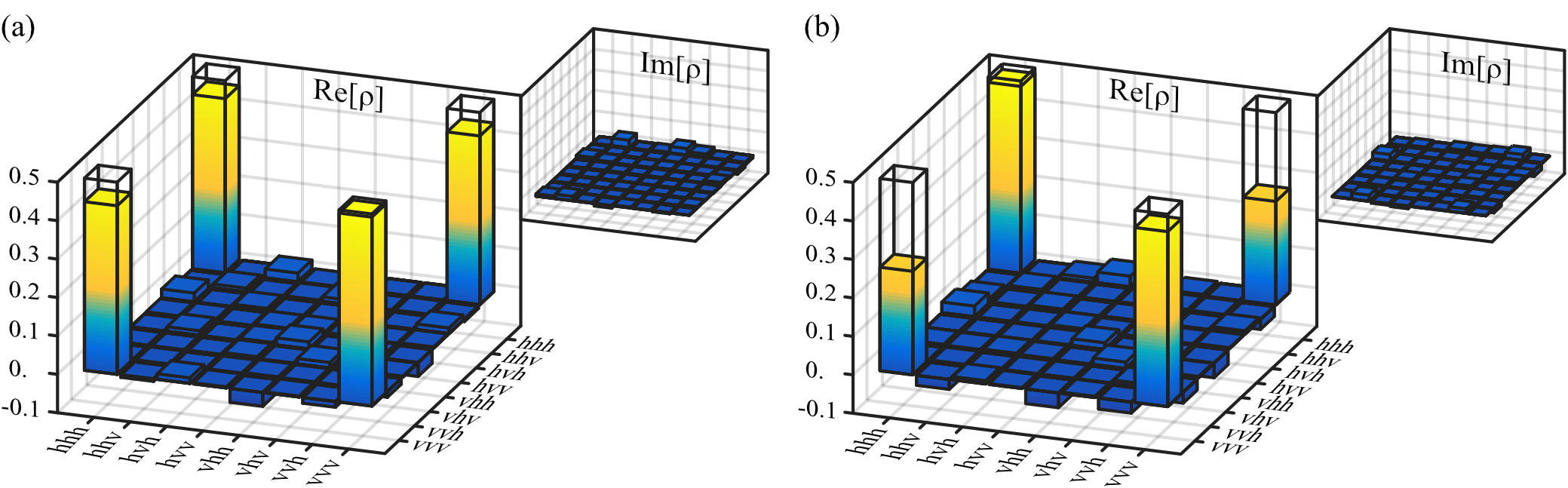}
    \caption{Reconstructed density matrices of the prepared GHZ states for (a) $\Delta = 0$~mm and (b) $\Delta =-3.5$~mm.}
    \label{fig:densityMatrices}
\end{figure}

\subsection{Tomography results}
In the experiment, the fourth photon is measured in the $X$ basis, and by post-selecting on the $|d\rangle\doteq(|h\rangle+|v\rangle)/\sqrt(2)$, we obtain the GHZ state,
\begin{equation}
    \rho = p \ket{GHZ}\bra{GHZ} + \frac{1-p}{2} (\ket{hhh}\bra{hhh} + \ket{vvv}\bra{vvv}),
\end{equation}
where $(1-p)$ is amount of controlled noise, which we use to test the protocol robustness, and is implemented via a relative temporal delay $\Delta$ added to one photon in the PBS gate.
With $\delta = 0$~mm we maximise $p$ and perform full QST to benchmark the initial quality of the state.
Since our polarisation analysers are equipped with single photon detection on both output ports of the PBS, we use the mutually-unbiased basis set, i.e., each photonic qubit is measured in the $Z$, $X$, and $Y$ basis.
In total, this required iterating through 27 measurement settings while recording the 8 outcomes simultaneously.
For each measurement setting we used an integration time of 30~seconds.
The dataset, consisting of 81 projective measurements, is used to reconstruct the density matrix using a maximum liklihood estimation algorithm.
At this setting we obtained the density matrix, $\rho$, shown in Fig~\ref{fig:densityMatrices}~(a).
From $\rho$ we evaluated the three-qubit GHZ state fidelity to be $0.923\pm0.005$ and a quantum state purity of $0.865\pm0.009$, errors are estimated via Monte Carlo simulation assuming Poissonian photon counting statistics.
For pure GHZ states both values can reach unity, in practice several sources of noise exist; multi-photon terms from the probabilistic sources, mode mismatch on the PBS gate, and finite polarisation extinction in the polarisation elements.
In our experiments we set the pump power to each EPS to be 100~mW, and an average four-fold coincidence rate of 15 per second was measured.
This pump power was chosen to balance the impact of multi-photon terms while ensuring the measurement times remain practical when ensuring for good statistics.

We also plot the density matrix for the largest delay offset in our experiment, $\Delta=-$3.5~mm in Fig~\ref{fig:densityMatrices}~(b).
In that configuration, we observe a fidelity to the three-qubit GHZ state of $0.73\pm0.02$ and a state purity of $0.58\pm0.02$.

\subsection{Local Phase Encoders}

\begin{figure}[h!]
    \centering
    \includegraphics[width=\linewidth]{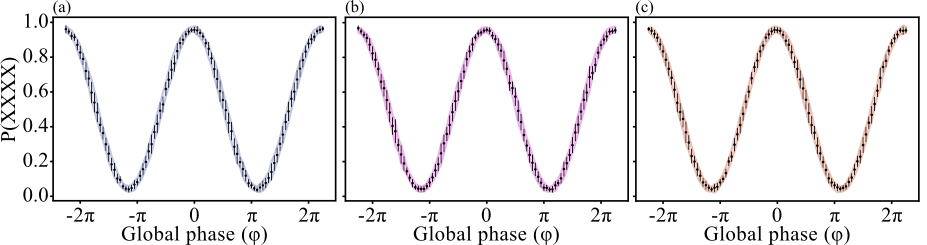}
    \caption{Measured visibilities of each sensor's phase encoding stage.
    In each case, two phase shifters are fixed in phase, while only one is rotated to sweep $[-2\pi, 2\pi]$.
    From the measured fringe patterns, we fit the data to estimate the visibilities of $91.5(3)\%$, $92.0(3)\%$ and $91.9(3)\%$ for (a), (b) and (c) respectively.
    For completeness we note that the fourth qubit is projected in X in the setup, hence the fourth qubit label.}
    \label{fig:phaseSweeps}
\end{figure}

For the three sensors, we set up the local phase shifters which consist of a QWP-HWP-QWP prior to the polarisation analysers.
Just like in the phase compensation in the PBS-G, the two QWPs are set to $+45^\circ$ from their optic axes, while the middle HWP is rotated to encode a nominal phase $\theta_i$.
To calibrate each local phase encoding device we set $\theta_i\in[-2\pi,2\pi]$ and measure in the joint-$X$ measurements which result in the fringe-like patterns in Fig.~\ref{fig:phaseSweeps}.
In the ideal case, each phase shifter should produce fringes with unit visibility, calculated as $v = \frac{p_{max} - p_{min}}{p_{max} + p_{min}}$, where $p_{max}$ and $p_{min}$ are the maximum and minimum measured probabilities.
For all three local phase shifters, we find the measured visibility (or contrast) in the joint-$X$ basis is $>0.90$.
We find that this deviation from unit visibility leads to the Cramer-Rao Bound featuring the phase dependence when estimating the average phase $\phi$ as shown in the manuscript.

\section{Analysis of privacy and integrity}
\label{sec:privacy_an_integrity}
\subsection{Privacy}
In the main text we introduced equation~(\ref{eq:verification-prob}) which relates the lower-bound fidelity of the output state $\tau$ to the GHZ state when performing the \textsc{State Verification} protocol with parameters {$n, f, c, m$}.
This bound was derived by relating the QFI to the state fidelity using the continuity equation\cite{augusiak2016asymptotic, rezakhani2019},
\begin{equation}
     | F_{Q|\theta} (\rho_\theta) - F_{Q|\theta} (\sigma_\theta)| \leq \xi \sqrt{1-F(\rho, \sigma)} N^2 \label{SM_cont}
\end{equation}
where $\xi$ is a constant which can vary depending on the rank of the density matrix. Substituting equation~(\ref{eq:verification-prob}) into (\ref{SM_cont}), we arrive at the expression for the private information derived in Ref.~\cite{Shettell2022arxiv, hassani2025},
\begin{equation}
    \varepsilon_p \leq  \frac{24}{n^2}\sqrt{\frac{2\sqrt{c}}{n} + 2nf},
    \label{eq:security}
\end{equation}
where $n$ is the number of sensors and $f$ is the failure rate of the stabilizer tests, averaged over all the $n+1$ stabilizers; conditional on the success from the \textsc{State Verification} protocol which must satisfy $f\leq 1/(2n^2)$.
The positive constant $c$ is a variable constrained by $3/(2m) < c < (n-1)^2/4$, which should ideally be minimised to decrease the upper-bounded $\varepsilon_p$ value.
Notably, the lower bound on $c$ is set by $m$ which directly relates to the number of test copies used, as captured by $N_t = \lceil 2mn^5\text{log}(n) \rceil$.
As such, to minimise $c$ we need more copies of the resource state during the verification step.
In the main body we assumed infinite copies ($N_t\rightarrow \infty$), thus $c \rightarrow 0$, which simplifies equation~\ref{eq:security} to $\varepsilon_p \leq 24 n^{-2} \sqrt{2nf}$.
The verifier can then choose the failure rate $f$ to accept in the \textsc{State Verification} protocol to obtain a satisfactory level of privacy and integrity in their protocol as this decreases $\varepsilon_p$.

In particular, $\varepsilon_p$-privacy quantifies the local sensor information leakage to an adversary.
We extend the privacy definition from Ref.~\cite{Shettell2022arxiv} to the case of $n$ parameters encoded on a multi-partite state via unitary encoding.
\begin{definition} \label{def:privacy}
    Let $\rho_{\theta_1,\dots,\theta_n}$ be a $n$-partite state encoded with $n$ parameters through the local unitaries $U_j=\exp(-i \theta_j H_j)$, where $H_j$ is a Hermitian operator on the $j$-th system. Then, the state is said to be $\varepsilon_p$-private if, $\forall\, j \in [1,n]$, there exists choices for the other local phases, $\phi_k(\theta_j) \mbox{ for } k\neq j$, such that: 
\begin{align}
   \quad F_{Q|\theta_j}(\rho_{\phi_1,\dots,\theta_j,\dots,\phi_n}) \leq \varepsilon_p (h_{\max} - h_{\min})^2, \label{privacy-def}
\end{align}
where $F_{Q|\theta_j}$ is the QFI calculated with respect to the parameter $\theta_j$ and where $h_{\max}$ ($h_{\min}$) is the largest (smallest) eigenvalue of the operator $\sum_{j=1}^n H_j$.
\end{definition}
Note that we expect $\varepsilon_p \in[0,1]$ since for the trivial choice $\phi_k=\theta_j$ we get $F_{Q|\theta_j}(\rho_{\theta_j,\dots,\theta_j}) \leq (h_{\max} - h_{\min})^2$ (this result follows from Theorem~1 in \cite{Fiderer2019}).

For the unitary encoding operated by the sensors in our experiment, we have $H_j=\sigma_Z/2$, which corresponds to the Definition~2 which recovers the definition provided in Ref.~\cite{Shettell2022arxiv}.

\begin{definition} \label{def:privacy-ourcase}
The $n$-partite state $\rho_{\theta_1,\dots,\theta_n}$, encoded by the local unitaries $U_j=\exp(-i \theta_j \sigma_Z/2)$, is said to be $\varepsilon_p$-private if, $\forall\, j \in [1,n]$, there exists $\phi_k(\theta_j) \mbox{ for } k\neq j$ such that: 
\begin{align}
   \quad F_{Q|\theta_j}(\rho_{\phi_1,\dots,\theta_j,\dots,\phi_n}) \leq \varepsilon_p n^2. \label{privacy-def-ourcase}
\end{align}
\end{definition}

\subsection{Integrity}

We now discuss the $\varepsilon_i$-integrity of the protocol. The authors of Ref.~\cite{Shettell2022arxiv} quantify the deviation in accuracy and in precision of the estimator of the protocol, when the shared resource deviates from a GHZ state. As such, these measures depend on an upper bound on the average trace distance between the encoded resource state and the encoded GHZ state---averaged over the parameter estimation rounds \cite{Shettell2022arxiv}. Thus, we have the following definition. 
\begin{definition} \label{def:integrity}
    Let $\tau_j$ be the target state of the $j$-th verification protocol after being encoded with the local phases, and let $T(\rho, \sigma)$ be the trace distance between $\rho$ and $\sigma$. Then, the metrology protocol is $\varepsilon_i$-integrous if:
\begin{align}
    \frac{1}{\nu} \sum_{j=1}^\nu T(\tau_j, \tau_{\phi}) \leq \varepsilon_i, \label{eq:def-epsilon-integrity}
\end{align}
where $\tau_{\phi}$ is the encoded GHZ state,
\begin{equation}
    \tau_\phi = \ket{\Phi}\bra{\Phi},\quad \ket{\Phi}=\frac{1}{\sqrt{2}}( | 0 \rangle^{\otimes n} + e^{i \phi}| 1 \rangle^{\otimes n} ).  \label{encodedGHZ}
\end{equation}
\end{definition}

The deviations in accuracy and precision derived in \cite{Shettell2022arxiv} are obtained for a specific estimator $\hat{\phi}$ of the global phase.
In particular, let $\tau_\phi$ be the ideal encoded state, \eqref{encodedGHZ}.
Then, the expectation value of $\sigma_X^{\otimes n}$ on $\tau_\phi$ is:
\begin{align}
    g(\phi) = \text{Tr}[\sigma_X^{\otimes n} \tau_\phi] = \cos\phi. \label{g(phi)}
\end{align}
On the other hand, the expectation value of $\sigma_X^{\otimes n}$ is approximated by the average of its measurement outcomes over the $\nu$ metrology rounds, when $\nu \gg 1$ and when the resource state is close to the GHZ state:
\begin{align}
    \hat{g} = \frac{1}{\nu} \sum_{j=1}^\nu m_j,
\end{align}
where $m_j$ is the outcome of $\sigma_X^{\otimes n}$ in the $j$-th metrology round (i.e., computed on $\tau_j$). The estimator employed in \cite{Shettell2022arxiv} is defined as:
\begin{align}
    \hat{\phi} &= g^{-1} (\hat{g}) \nonumber\\
    &= \arccos(\hat{g}). \label{estimator}
\end{align}
The authors in \cite{Shettell2022arxiv} argue that the estimator in \eqref{estimator}, for $\nu \gg 1$, is unbiased when the resource state coincides with the ideal state ($\tau_j =\tau_\phi$ for all $j$). Then, Theorem~1 in \cite{Shettell2022arxiv} provides a bound on the bias of the estimator computed on the actual resource state:
\begin{equation}\label{eq:accuracy}
    |\mathbb{E}(\hat{\phi}) - \phi| \leq \frac{2o\varepsilon_i}{|\sin \phi|},
\end{equation}
where $\mathbb{E}(\hat{\phi})$ is the expectation value of \eqref{estimator} on the actual resource state, while $o$ is the maximum magnitude of the eigenvalues of the observable which is being measured for parameter estimation; in our case $o=1$.

Similarly, one can argue that the variance of the estimator in \eqref{estimator}, assuming that the resource state is the same in every parameter estimation round ($\tau_j = \tau$), can be computed as \cite{Shettell2022arxiv}:
\begin{align}
    \Delta^2 \hat{\phi} = \frac{1-(\text{Tr}[\sigma_X^{\otimes n} \tau])^2}{\nu \, \sin^2\phi }. \label{variance-formula}
\end{align}
Now, when the resource state is the encoded GHZ state, the variance simplifies to $1/\nu$. Theorem~1 in \cite{Shettell2022arxiv} provides a bound on the deviation of the precision of the estimator when the resource state is not ideal:
\begin{equation}\label{eq:precision}
    \left\lvert \Delta^2 \hat{\phi} - \frac{1}{\nu}\right\rvert \leq \frac{4 \varepsilon_i (2\nu^{-1} + \varepsilon_i)}{\sin^2 \phi}.
\end{equation}
We evaluate the bounds in \eqref{eq:precision} and \eqref{eq:accuracy} by replacing $\mathbb{E}(\hat{\phi})$ with $\hat{\phi}$ and $\Delta^2 \hat{\phi}$ with the formula in \eqref{variance-formula}, where we choose $\tau=\sum_j \tau_j/\nu$:
\begin{align}
    \Delta^2 \hat{\phi} &= \frac{1-(\mathbb{E}(\hat{g}))^2}{\nu \, \sin^2\phi } \\
    &\approx \frac{1-(\hat{g})^2}{\nu \, \sin^2\phi }.
\end{align}

\subsection{Experimental validation of integrity}

The $\varepsilon_i$ parameter is the average trace distance between the target state output by the verification protocol and the ideal GHZ state.
This can be calculated through one of two ways, either taken directly from the experiment or from the theoretical framework outlined in Ref.~\cite{Shettell2022arxiv}. 
We will show that the experimentally obtained $\varepsilon_i$ term leads to a tighter bound than the current theoretical term. 
First, using the experimental fidelity achieved, the upper bound on the trace distance via the fidelity is $\textrm{Tr}(\rho-\sigma)\leq\sqrt{1-F(\rho,\sigma)}$, where $\varepsilon_{i,\textrm{exp}}=\sqrt{1-F(\rho,\sigma)}$.
The second method relies exclusively on the information provided by the protocol in \cite{Shettell2022arxiv}, which states that after the verification protocol the fidelity must be above a certain threshold Eq.~\eqref{eq:verification-prob} in the main text. 
Using this lower bound and the trace distance via the fidelity, we get $\varepsilon_{i,\textrm{theo}}=\sqrt{2\sqrt{c}/n+2nf}$. 
We can now compare these two integrity metrics, using one of the phase sweep plots presented in Fig.~\ref{fig:phaseSweeps}. 

\begin{figure}[t]
    \centering
    \includegraphics[width=0.8\linewidth]{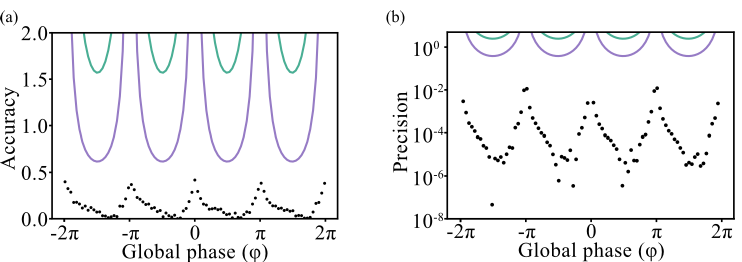}
    \caption{The accuracy and precision of the protocol, defined with respect to $\varepsilon_i$. In both plots, the green line is defined w.r.t $\varepsilon_{i,\textrm{theo}}=\sqrt{2\sqrt{c} / n+2nf}$ and the purple line is defined w.r.t $\varepsilon_{i,\textrm{exp}}=\sqrt{1-F(\rho,\sigma)}$. (a) shows the accuracy, as in \eqref{eq:accuracy}, of the measured global phase (black line). Specifically, the green and purple are the right hand side of the inequality and the black line is the left hand side of the inequality in \eqref{eq:accuracy}. (b) shows the precision, a more direct measure of integrity as in \eqref{eq:precision} it is w.r.t a certain number of parameter estimation rounds $\nu$. The green and purple lines are the right hand side and the black line is the left hand side of the inequality in \eqref{eq:precision}. This is taken at $\nu = 3200$, $c = 0.25$, with $f = 0.047$ as measured in \eqref{eq:accuracy}, and the fidelity used is 0.905, the lower bound of 0.907(2).}
    \label{fig:integrity_supp}
\end{figure}

The accuracy and precision and hence the integrity of the sensors can be plotted, as presented in Fig.\ref{fig:integrity_supp}. 
The accuracy is indeed loose in both regards of $\varepsilon_{i,\textrm{exp}}$ and $\varepsilon_{i,\textrm{theo}}$, however our experimentally obtained integrity
parameter is substantially tighter than the current theoretical framework. 
The measured precision appears to be up to seven orders of magnitude away from the theoretical bound, in both the $\varepsilon_{i,\textrm{exp}}$ and $\varepsilon_{i,\textrm{theo}}$ expressions. It should be noted
that for $\varepsilon_{i,\textrm{exp}}\rightarrow 0$, the theoretical expression scales exponentially towards our precision. 
This hints that modeling $\varepsilon_{i,\textrm{exp}}$ through the trace distance measure leads to a loose criteria,
but not as loose as previously established through $\varepsilon_{i,\textrm{theo}}$, which does not get close to the characteristics determined from the experimental data—-thus concluding and quantifying the looseness.
We have therefore provided an alternative, tighter, bound in both the accuracy and precision of the protocol, leading to the integrity parameter being more honest.

\section{Direct calculation of QFI for privacy statement}
\label{sec:QFIcalc}

In this section we perform a direct calculation of the QFI appearing in the privacy definition that applies to our experiment, Definition~\ref{def:privacy-ourcase}.
This is achieved by employing the tomographically-reconstructed resource state of our experiment.
The goal is to compare the direct calculation of the QFI, which would lead to a tight privacy bound $\varepsilon_p$, with the upper bound on $\varepsilon_p$ provided by \cite{Shettell2022arxiv} and reported in \eqref{eq:security}.

We start by recalling that the QFI of a state $\rho_{\theta}$, obtained by encoding $\theta$ into $\rho$, is given by \cite{ParisQFI}:
\begin{align}
    F_{Q|\theta}(\rho_\theta) = 2 \sum_{k,l} \frac{\left\lvert \braket{\psi_k|\partial_\theta \rho_\theta|\psi_l}\right\rvert^2}{\lambda_k + \lambda_l} , \label{QFI}
\end{align}
where the sum only runs over indices for which $\lambda_k + \lambda_l \neq 0$ and where $\lambda_k$ and $\ket{\psi_k}$ are obtained from the spectral decomposition of $\rho_\theta$: $\rho_\theta=\sum_k \lambda_k \ket{\psi_k}\bra{\psi_k}$ (where $\lambda_k$ can also be zero). When the parameter $\theta$ is encoded by the unitary transformation:
\begin{align}
    \rho_\theta = e^{-i \theta A} \rho e^{i \theta A}, \label{unitary-encoding}
\end{align}
one can show that the QFI reduces to:
\begin{align}
    F_{Q|\theta}(\rho_\theta) = 2 \sum_{k,l} \frac{(\lambda_k - \lambda_l)^2}{\lambda_k + \lambda_l} \left\lvert \braket{\varphi_k|A|\varphi_l}\right\rvert^2, \label{QFI-unitary-encoding}
\end{align}
where now $\lambda_k$ and $\ket{\varphi_k}$ are obtained from the spectral decomposition of the unencoded state $\rho$: $\rho = \sum_k \lambda_k \ket{\varphi_k}\bra{\varphi_k} $, implying that the QFI in \eqref{QFI-unitary-encoding} is independent of the value of $\theta$. Moreover, if the state $\rho$ is pure, $\rho=\ket{\varphi}\bra{\varphi}$, it holds:
\begin{align}
    F_{Q|\theta}(\rho_\theta) = 4 \left[\braket{\varphi|A^2 | \varphi} - \left(\braket{\varphi|A|\varphi}\right)^2 \right].
\end{align}

In order to determine the privacy parameter $\varepsilon_p$, let us fix for the moment $j=1$ in \eqref{privacy-def-ourcase}, implying that we want to make a statement about the privacy of the local phase $\theta_1$ when encoded in  our resource state $\rho$. In order to find the best possible privacy parameter $\varepsilon_1$, we solve the following optimization problem:
\begin{align}
    \alpha_1=\min_{\phi_k(\theta), \, k \neq 1} F_{Q|\theta_1}(\rho_{\theta_1,\phi_2(\theta_1),\dots,\phi_n(\theta_1)}), \label{optimization}
\end{align}
and then choose $\varepsilon_1=\alpha_1/n^2$. The optimization is done over all possible choices of functions $\phi_k(\theta)$, such that the resulting state is functionally independent of $\theta_1$ --or as independent as possible. If we repeat this procedure for each $j$, we can then set $\varepsilon=\max_j \varepsilon_j$ and claim that the resource state $\rho$ is $\varepsilon_p$-private according to definition \eqref{privacy-def-ourcase}. To solve the optimization in \eqref{optimization}, we compute the QFI, with respect to $\theta_1$, of the encoded resource state:
\begin{align}
    \rho_{\theta_1,\phi_2(\theta_1),\dots,\phi_n(\theta_1)} = e^{-i \theta_1 \frac{Z_1}{2}} e^{-i \phi_2(\theta_1) \frac{Z_2}{2}} \cdots e^{-i \phi_n(\theta_1) \frac{Z_n}{2}} \, \rho \,   e^{i \theta_1 \frac{Z_1}{2}} e^{i \phi_2(\theta_1) \frac{Z_2}{2}} \cdots e^{i \phi_n(\theta_1) \frac{Z_n}{2}}, \label{state-func-theta1}
\end{align}
where $\rho$ is the resource state outputted by the verification procedure. To this aim, we start from the QFI formula in \eqref{QFI} and derive a generalized version of \eqref{QFI-unitary-encoding} which is valid when the state is encoded with multiple parameters via unitary encoding:
\begin{align}
    \rho_\theta = e^{-i \theta A_1} e^{-i \phi_2(\theta) A_2} \cdots e^{-i \phi_n(\theta) A_n} \, \rho \,   e^{i \theta A_1} e^{i \phi_2(\theta) A_2} \cdots e^{i \phi_n(\theta) A_n}, \label{state-func-theta}
\end{align}
with the additional assumption that the generators $A_j$ commute pairwise. Then, the encoded resource state in \eqref{state-func-theta1} is a particular case of \eqref{state-func-theta}. We start by computing the derivative of the state in \eqref{state-func-theta} with respect to $\theta$:
\begin{align}
    \partial_\theta \rho_\theta = U_\theta [-i A_1,\rho] U^\dag_\theta + \sum_{k=2}^n  U_\theta [-i A_k \phi'_k,\rho] U^\dag_\theta,
\end{align}
where we introduced a short-hand notation for the unitary encoding:
\begin{align}
    U_\theta := e^{-i \theta A_1} e^{-i \phi_2(\theta) A_2} \cdots e^{-i \phi_n(\theta) A_n}.
\end{align}
Then, we observe that the spectral decomposition of $\rho_\theta$, in the case of unitary encoding, is obtained from that of $\rho$; namely, the eigenvalues are unchanged ($\lambda_l$) and the eigenvectors of $\rho_\theta$ are given by: $\ket{\psi_l}=U_\theta \ket{\varphi_l}$, with $\ket{\varphi_l}$ the eigenvectors of $\rho$. This implies that:
\begin{align}
    \braket{\psi_m|\partial_\theta \rho_\theta|\psi_l} &= \bra{\varphi_m} \left([-i A_1,\rho] + \sum_{k=2}^n  [-i A_k \phi'_k,\rho] \right) \ket{\varphi_l} \nonumber\\
    &= i (\lambda_m - \lambda_l) \left(\braket{\varphi_m|A_1|\varphi_l} + \sum_{k=2}^n \phi'_k (\theta) \braket{\varphi_m|A_k|\varphi_l}\right),
\end{align}
which substituted in \eqref{QFI} yields:
\begin{align}
    F_{Q|\theta}(\rho_\theta) = 2 \sum_{m,l} \frac{(\lambda_m - \lambda_l)^2}{\lambda_m + \lambda_l} \left\lvert \braket{\varphi_m|A_1|\varphi_l} + \sum_{k=2}^n \phi'_k (\theta) \braket{\varphi_m|A_j|\varphi_l} \right\rvert^2, \label{generalized-QFI-unitary-encoding}
\end{align}
which is the QFI of the state \eqref{state-func-theta}, with commuting generators and where $\lambda_l$ and $\ket{\varphi_l}$ are obtained from the spectral decomposition of the unencoded state: $\rho=\sum_l \lambda_l \ket{\varphi_l}\bra{\varphi_l}$. Now, we apply the last expression for the state in \eqref{state-func-theta1} to solve the optimization problem in \eqref{optimization}. We obtain:
\begin{align}
    \alpha_1=\frac{1}{2} \min_{\phi_k(\theta), \, k \neq 1}  \sum_{m,l} \frac{(\lambda_m - \lambda_l)^2}{\lambda_m + \lambda_l} \left\lvert \braket{\varphi_m|Z_1|\varphi_l} + \sum_{k=2}^n \phi'_k (\theta) \braket{\varphi_m|Z_k|\varphi_l} \right\rvert^2, \label{optimization2}
\end{align}
where the solution of the optimization, $\alpha_1$, can potentially depend on $\theta$ if the derivatives of the functions $\phi_k (\theta)$ are not constant. However, we can assume the functions $\phi_k (\theta)$ to be of the form: $\phi_k (\theta)= c_k \theta$, for $c_k \in \mathbbm{R}$, without loss of generality. Indeed, suppose that $\tilde{\phi}_k(\theta)$ are the optimal functions that solve the optimization problem, and suppose that $\alpha_1(\theta)$  depends on $\theta$ through $\tilde{\phi}'_k(\theta)$. Let $\bar{\theta}$ be the parameter that minimizes $\alpha_1$, i.e.: $\bar{\theta} = \arg \min_\theta \alpha_1 (\theta)$. Then, we can choose another set of functions directly proportional to $\theta$, namely  $\phi_k (\theta)=\theta \tilde{\phi}'_k(\bar{\theta})$, such that the resulting QFI is minimal and equal to $\alpha_1 (\bar{\theta})$, with the added benefit of being independent of $\theta$. Therefore, we can restrict the optimization in \eqref{optimization2} to linear functions of $\theta$. Thus, we get the following parameter for the privacy of $\theta_1$ when encoded in the resource state:
\begin{align}
    \varepsilon_1=\frac{1}{2 n^2} \min_{c_k \in \mathbbm{R}}  \sum_{m,l} \frac{(\lambda_m - \lambda_l)^2}{\lambda_m + \lambda_l} \left\lvert \braket{\varphi_m|Z_1|\varphi_l} + \sum_{k=2}^n c_k  \braket{\varphi_m|Z_k|\varphi_l} \right\rvert^2. \label{optimization3}
\end{align}
By repeating this procedure for each $j$ and by taking the maximum of the privacy parameters, we obtain the desired $\varepsilon_p$ parameter for our resource state.

Hence, the resource state $\rho$, with spectral decomposition $\rho=\sum_l \lambda_l \ket{\varphi_l}\bra{\varphi_l}$, is $\varepsilon_p$-private (according to Definition~\ref{def:privacy}) with respect to the parameters $\theta_1 ,\dots, \theta_n$ when encoded by $U_j=e^{-i \theta_j \frac{{Z_j}}{2}}$, where:
\begin{align}
    \varepsilon_p = \frac{1}{2 n^2} \max_{1 \leq j \leq n} \,\, \min_{c_{k|j} \in \mathbbm{R}}  \sum_{m,l} \frac{(\lambda_m - \lambda_l)^2}{\lambda_m + \lambda_l} \left\lvert \braket{\varphi_m|Z_j|\varphi_l} + \sum_{k\neq j} c_{k|j}   \braket{\varphi_m|Z_k|\varphi_l} \right\rvert^2.  \label{privacy-param}
\end{align}

\noindent\textbf{Remark:} We verify that the GHZ state encoded by $U_j=e^{-i \theta_j \frac{{Z_j}}{2}}$, for $j=1,\dots,n$, is $0$-private. A simple argument follows directly from Definition~\ref{def:privacy}. Indeed, one can choose $\phi_k(\theta_j)=-\theta_j/(n-1)$, for $k\neq j$, and observe that the resulting encoded state is independent of $\theta_j$:
\begin{align}
    \rho_{\phi_1,\dots,\theta_j,\dots,\phi_n}= U_1 U_2 \cdots U_n \ket{GHZ}\bra{GHZ} U^\dag_1 U^\dag_2 \cdots U^\dag_n = \ket{GHZ}\bra{GHZ}.
\end{align}
Thus, by definition of the QFI in \eqref{QFI}, it holds: $F_{Q|\theta_j}(\rho_{\phi_1,\dots,\theta_j,\dots,\phi_n})=0$ for every $j$ and hence $\varepsilon_p=0$ according to Definition~\ref{def:privacy}. Alternatively, we can directly evaluate the privacy parameter from \eqref{privacy-param}. For this, we start from the spectral decomposition of the GHZ state (where we identified $\ket{\varphi_1}=\ket{GHZ}$):
\begin{align}
    \rho = 1 \cdot \ket{GHZ}\bra{GHZ} + 0 \cdot(\mathbbm{1} - \ket{GHZ}\bra{GHZ}),
\end{align}
which employed in \eqref{privacy-param} yields:
\begin{align}
    \varepsilon_p &= \frac{1}{ n^2} \max_{1 \leq j \leq n} \,\, \min_{c_{k|j} \in \mathbbm{R}}  \sum_{m>1} \left\lvert \braket{\varphi_m|Z_j|GHZ} + \sum_{k\neq j} c_{k|j}   \braket{\varphi_m|Z_k|GHZ} \right\rvert^2 \nonumber\\
    &=  \frac{1}{ n^2} \max_{1 \leq j \leq n} \,\, \min_{c_{k|j} \in \mathbbm{R}}  \sum_{m>1} \left\lvert \bra{\varphi_m}\left( Z_j  + \sum_{k\neq j} c_{k|j} Z_k \right)  \ket{GHZ} \right\rvert^2 \nonumber\\
    &=  \frac{1}{ n^2} \max_{1 \leq j \leq n} \,\, \min_{c_{k|j} \in \mathbbm{R}}  \sum_{m>1}   \bra{GHZ}\left( Z_j  + \sum_{k\neq j} c_{k|j} Z_k \right)^\dag  \ket{\varphi_m} \bra{\varphi_m}\left( Z_j  + \sum_{k\neq j} c_{k|j} Z_k \right)  \ket{GHZ} \nonumber\\
    &=  \frac{1}{ n^2} \max_{1 \leq j \leq n} \,\, \min_{c_{k|j} \in \mathbbm{R}}    \bra{GHZ}\left( Z_j  + \sum_{k\neq j} c_{k|j} Z_k \right)^\dag  \left(\mathbbm{1} - \ket{GHZ}\bra{GHZ} \right)\left( Z_j  + \sum_{k\neq j} c_{k|j} Z_k \right)  \ket{GHZ} \nonumber\\
    &=  \frac{1}{ n^2} \max_{1 \leq j \leq n} \,\, \min_{c_{k|j} \in \mathbbm{R}}    \left[\braket{GHZ|(Z'_j)^2|GHZ} -  \left(\braket{GHZ|Z'_j|GHZ}\right)^2 \right],
\end{align}
where we defined:
\begin{align}
    Z'_j= Z_j  + \sum_{k\neq j} c_{k|j} Z_k.
\end{align}
Now, due to the structure of the GHZ state, we can e.g. choose $c_{k|j}=-1/(n-1)$, such that $Z'_j \ket{GHZ} = 0$, for every $j$. This yields $\varepsilon_p=0$, as claimed.


\begin{thebibliography}{50}%
\makeatletter
\providecommand \@ifxundefined [1]{%
 \@ifx{#1\undefined}
}%
\providecommand \@ifnum [1]{%
 \ifnum #1\expandafter \@firstoftwo
 \else \expandafter \@secondoftwo
 \fi
}%
\providecommand \@ifx [1]{%
 \ifx #1\expandafter \@firstoftwo
 \else \expandafter \@secondoftwo
 \fi
}%
\providecommand \natexlab [1]{#1}%
\providecommand \enquote  [1]{``#1''}%
\providecommand \bibnamefont  [1]{#1}%
\providecommand \bibfnamefont [1]{#1}%
\providecommand \citenamefont [1]{#1}%
\providecommand \href@noop [0]{\@secondoftwo}%
\providecommand \href [0]{\begingroup \@sanitize@url \@href}%
\providecommand \@href[1]{\@@startlink{#1}\@@href}%
\providecommand \@@href[1]{\endgroup#1\@@endlink}%
\providecommand \@sanitize@url [0]{\catcode `\\12\catcode `\$12\catcode `\&12\catcode `\#12\catcode `\^12\catcode `\_12\catcode `\%12\relax}%
\providecommand \@@startlink[1]{}%
\providecommand \@@endlink[0]{}%
\providecommand \url  [0]{\begingroup\@sanitize@url \@url }%
\providecommand \@url [1]{\endgroup\@href {#1}{\urlprefix }}%
\providecommand \urlprefix  [0]{URL }%
\providecommand \Eprint [0]{\href }%
\providecommand \doibase [0]{http://dx.doi.org/}%
\providecommand \selectlanguage [0]{\@gobble}%
\providecommand \bibinfo  [0]{\@secondoftwo}%
\providecommand \bibfield  [0]{\@secondoftwo}%
\providecommand \translation [1]{[#1]}%
\providecommand \BibitemOpen [0]{}%
\providecommand \bibitemStop [0]{}%
\providecommand \bibitemNoStop [0]{.\EOS\space}%
\providecommand \EOS [0]{\spacefactor3000\relax}%
\providecommand \BibitemShut  [1]{\csname bibitem#1\endcsname}%
\let\auto@bib@innerbib\@empty
\bibitem [{\citenamefont {Gisin}\ \emph {et~al.}(2002)\citenamefont {Gisin}, \citenamefont {Ribordy}, \citenamefont {Tittel},\ and\ \citenamefont {Zbinden}}]{Gisin2002}%
  \BibitemOpen
  \bibfield  {author} {\bibinfo {author} {\bibfnamefont {N.}~\bibnamefont {Gisin}}, \bibinfo {author} {\bibfnamefont {G.}~\bibnamefont {Ribordy}}, \bibinfo {author} {\bibfnamefont {W.}~\bibnamefont {Tittel}}, \ and\ \bibinfo {author} {\bibfnamefont {H.}~\bibnamefont {Zbinden}},\ }\href@noop {} {\bibfield  {journal} {\bibinfo  {journal} {Review of Modern Physics}\ }\textbf {\bibinfo {volume} {74}},\ \bibinfo {pages} {145} (\bibinfo {year} {2002})}\BibitemShut {NoStop}%
\bibitem [{\citenamefont {Pirandola}\ \emph {et~al.}(2020)\citenamefont {Pirandola}, \citenamefont {Andersen}, \citenamefont {Banchi}, \citenamefont {Berta}, \citenamefont {Bunandar}, \citenamefont {Colbeck}, \citenamefont {Englund}, \citenamefont {Gehring}, \citenamefont {Lupo}, \citenamefont {Ottaviani}, \citenamefont {Pereira}, \citenamefont {Razavi}, \citenamefont {Shaari}, \citenamefont {Tomamichel}, \citenamefont {Usenko}, \citenamefont {Vallone}, \citenamefont {Villoresi},\ and\ \citenamefont {Wallden}}]{Pirandola2020}%
  \BibitemOpen
  \bibfield  {author} {\bibinfo {author} {\bibfnamefont {S.}~\bibnamefont {Pirandola}}, \bibinfo {author} {\bibfnamefont {U.~L.}\ \bibnamefont {Andersen}}, \bibinfo {author} {\bibfnamefont {L.}~\bibnamefont {Banchi}}, \bibinfo {author} {\bibfnamefont {M.}~\bibnamefont {Berta}}, \bibinfo {author} {\bibfnamefont {D.}~\bibnamefont {Bunandar}}, \bibinfo {author} {\bibfnamefont {R.}~\bibnamefont {Colbeck}}, \bibinfo {author} {\bibfnamefont {D.}~\bibnamefont {Englund}}, \bibinfo {author} {\bibfnamefont {T.}~\bibnamefont {Gehring}}, \bibinfo {author} {\bibfnamefont {C.}~\bibnamefont {Lupo}}, \bibinfo {author} {\bibfnamefont {C.}~\bibnamefont {Ottaviani}}, \bibinfo {author} {\bibfnamefont {J.~L.}\ \bibnamefont {Pereira}}, \bibinfo {author} {\bibfnamefont {M.}~\bibnamefont {Razavi}}, \bibinfo {author} {\bibfnamefont {J.~S.}\ \bibnamefont {Shaari}}, \bibinfo {author} {\bibfnamefont {M.}~\bibnamefont {Tomamichel}}, \bibinfo {author} {\bibfnamefont {V.~C.}\ \bibnamefont {Usenko}}, \bibinfo {author} {\bibfnamefont
  {G.}~\bibnamefont {Vallone}}, \bibinfo {author} {\bibfnamefont {P.}~\bibnamefont {Villoresi}}, \ and\ \bibinfo {author} {\bibfnamefont {P.}~\bibnamefont {Wallden}},\ }\href@noop {} {\bibfield  {journal} {\bibinfo  {journal} {Advances in Optics and Photonics}\ }\textbf {\bibinfo {volume} {12}},\ \bibinfo {pages} {1012} (\bibinfo {year} {2020})}\BibitemShut {NoStop}%
\bibitem [{\citenamefont {Barz}\ \emph {et~al.}(2012)\citenamefont {Barz}, \citenamefont {Kashefi}, \citenamefont {Broadbent}, \citenamefont {Fitzsimons}, \citenamefont {Zeilinger},\ and\ \citenamefont {Walther}}]{barz2012demonstration}%
  \BibitemOpen
  \bibfield  {author} {\bibinfo {author} {\bibfnamefont {S.}~\bibnamefont {Barz}}, \bibinfo {author} {\bibfnamefont {E.}~\bibnamefont {Kashefi}}, \bibinfo {author} {\bibfnamefont {A.}~\bibnamefont {Broadbent}}, \bibinfo {author} {\bibfnamefont {J.~F.}\ \bibnamefont {Fitzsimons}}, \bibinfo {author} {\bibfnamefont {A.}~\bibnamefont {Zeilinger}}, \ and\ \bibinfo {author} {\bibfnamefont {P.}~\bibnamefont {Walther}},\ }\href@noop {} {\bibfield  {journal} {\bibinfo  {journal} {Science}\ }\textbf {\bibinfo {volume} {335}},\ \bibinfo {pages} {303} (\bibinfo {year} {2012})}\BibitemShut {NoStop}%
\bibitem [{\citenamefont {Fitzsimons}(2017)}]{fitzsimons2017private}%
  \BibitemOpen
  \bibfield  {author} {\bibinfo {author} {\bibfnamefont {J.~F.}\ \bibnamefont {Fitzsimons}},\ }\href@noop {} {\bibfield  {journal} {\bibinfo  {journal} {npj Quantum Information}\ }\textbf {\bibinfo {volume} {3}},\ \bibinfo {pages} {23} (\bibinfo {year} {2017})}\BibitemShut {NoStop}%
\bibitem [{\citenamefont {Gottesman}\ \emph {et~al.}(2012)\citenamefont {Gottesman}, \citenamefont {Jennewein},\ and\ \citenamefont {Croke}}]{gottesman2012telescope}%
  \BibitemOpen
  \bibfield  {author} {\bibinfo {author} {\bibfnamefont {D.}~\bibnamefont {Gottesman}}, \bibinfo {author} {\bibfnamefont {T.}~\bibnamefont {Jennewein}}, \ and\ \bibinfo {author} {\bibfnamefont {S.}~\bibnamefont {Croke}},\ }\href@noop {} {\bibfield  {journal} {\bibinfo  {journal} {Physical Review Letters}\ }\textbf {\bibinfo {volume} {109}},\ \bibinfo {pages} {070503} (\bibinfo {year} {2012})}\BibitemShut {NoStop}%
\bibitem [{\citenamefont {Komar}\ \emph {et~al.}(2014)\citenamefont {Komar}, \citenamefont {Kessler}, \citenamefont {Bishof}, \citenamefont {Jiang}, \citenamefont {S{\o}rensen}, \citenamefont {Ye},\ and\ \citenamefont {Lukin}}]{Komar2014clocks}%
  \BibitemOpen
  \bibfield  {author} {\bibinfo {author} {\bibfnamefont {P.}~\bibnamefont {Komar}}, \bibinfo {author} {\bibfnamefont {E.~M.}\ \bibnamefont {Kessler}}, \bibinfo {author} {\bibfnamefont {M.}~\bibnamefont {Bishof}}, \bibinfo {author} {\bibfnamefont {L.}~\bibnamefont {Jiang}}, \bibinfo {author} {\bibfnamefont {A.~S.}\ \bibnamefont {S{\o}rensen}}, \bibinfo {author} {\bibfnamefont {J.}~\bibnamefont {Ye}}, \ and\ \bibinfo {author} {\bibfnamefont {M.~D.}\ \bibnamefont {Lukin}},\ }\href@noop {} {\bibfield  {journal} {\bibinfo  {journal} {Nature Physics}\ }\textbf {\bibinfo {volume} {10}},\ \bibinfo {pages} {582} (\bibinfo {year} {2014})}\BibitemShut {NoStop}%
\bibitem [{\citenamefont {Baumgratz}\ and\ \citenamefont {Datta}(2016)}]{baumgratz2016quantum}%
  \BibitemOpen
  \bibfield  {author} {\bibinfo {author} {\bibfnamefont {T.}~\bibnamefont {Baumgratz}}\ and\ \bibinfo {author} {\bibfnamefont {A.}~\bibnamefont {Datta}},\ }\href@noop {} {\bibfield  {journal} {\bibinfo  {journal} {Physical Review Letters}\ }\textbf {\bibinfo {volume} {116}},\ \bibinfo {pages} {030801} (\bibinfo {year} {2016})}\BibitemShut {NoStop}%
\bibitem [{\citenamefont {Liu}\ \emph {et~al.}(2021)\citenamefont {Liu}, \citenamefont {Zhang}, \citenamefont {Li}, \citenamefont {Zhang}, \citenamefont {Yin}, \citenamefont {Fei}, \citenamefont {Li}, \citenamefont {Liu}, \citenamefont {Xu}, \citenamefont {Chen} \emph {et~al.}}]{Liu2021}%
  \BibitemOpen
  \bibfield  {author} {\bibinfo {author} {\bibfnamefont {L.-Z.}\ \bibnamefont {Liu}}, \bibinfo {author} {\bibfnamefont {Y.-Z.}\ \bibnamefont {Zhang}}, \bibinfo {author} {\bibfnamefont {Z.-D.}\ \bibnamefont {Li}}, \bibinfo {author} {\bibfnamefont {R.}~\bibnamefont {Zhang}}, \bibinfo {author} {\bibfnamefont {X.-F.}\ \bibnamefont {Yin}}, \bibinfo {author} {\bibfnamefont {Y.-Y.}\ \bibnamefont {Fei}}, \bibinfo {author} {\bibfnamefont {L.}~\bibnamefont {Li}}, \bibinfo {author} {\bibfnamefont {N.-L.}\ \bibnamefont {Liu}}, \bibinfo {author} {\bibfnamefont {F.}~\bibnamefont {Xu}}, \bibinfo {author} {\bibfnamefont {Y.-A.}\ \bibnamefont {Chen}},  \emph {et~al.},\ }\href@noop {} {\bibfield  {journal} {\bibinfo  {journal} {Nature Photonics}\ }\textbf {\bibinfo {volume} {15}},\ \bibinfo {pages} {137} (\bibinfo {year} {2021})}\BibitemShut {NoStop}%
\bibitem [{\citenamefont {Nichol}\ \emph {et~al.}(2022)\citenamefont {Nichol}, \citenamefont {Srinivas}, \citenamefont {Nadlinger}, \citenamefont {Drmota}, \citenamefont {Main}, \citenamefont {Araneda}, \citenamefont {Ballance},\ and\ \citenamefont {Lucas}}]{nichol2022clocks}%
  \BibitemOpen
  \bibfield  {author} {\bibinfo {author} {\bibfnamefont {B.~C.}\ \bibnamefont {Nichol}}, \bibinfo {author} {\bibfnamefont {R.}~\bibnamefont {Srinivas}}, \bibinfo {author} {\bibfnamefont {D.}~\bibnamefont {Nadlinger}}, \bibinfo {author} {\bibfnamefont {P.}~\bibnamefont {Drmota}}, \bibinfo {author} {\bibfnamefont {D.}~\bibnamefont {Main}}, \bibinfo {author} {\bibfnamefont {G.}~\bibnamefont {Araneda}}, \bibinfo {author} {\bibfnamefont {C.}~\bibnamefont {Ballance}}, \ and\ \bibinfo {author} {\bibfnamefont {D.}~\bibnamefont {Lucas}},\ }\href@noop {} {\bibfield  {journal} {\bibinfo  {journal} {Nature}\ }\textbf {\bibinfo {volume} {609}},\ \bibinfo {pages} {689} (\bibinfo {year} {2022})}\BibitemShut {NoStop}%
\bibitem [{\citenamefont {Broadbent}\ \emph {et~al.}(2009)\citenamefont {Broadbent}, \citenamefont {Fitzsimons},\ and\ \citenamefont {Kashefi}}]{Broadbent2009}%
  \BibitemOpen
  \bibfield  {author} {\bibinfo {author} {\bibfnamefont {A.}~\bibnamefont {Broadbent}}, \bibinfo {author} {\bibfnamefont {J.}~\bibnamefont {Fitzsimons}}, \ and\ \bibinfo {author} {\bibfnamefont {E.}~\bibnamefont {Kashefi}},\ }in\ \href@noop {} {\emph {\bibinfo {booktitle} {2009 50th annual IEEE symposium on foundations of computer science}}}\ (\bibinfo {organization} {IEEE},\ \bibinfo {year} {2009})\ pp.\ \bibinfo {pages} {517--526}\BibitemShut {NoStop}%
\bibitem [{\citenamefont {Hahn}\ \emph {et~al.}(2020)\citenamefont {Hahn}, \citenamefont {de~Jong},\ and\ \citenamefont {Pappa}}]{hahn2020anonymous}%
  \BibitemOpen
  \bibfield  {author} {\bibinfo {author} {\bibfnamefont {F.}~\bibnamefont {Hahn}}, \bibinfo {author} {\bibfnamefont {J.}~\bibnamefont {de~Jong}}, \ and\ \bibinfo {author} {\bibfnamefont {A.}~\bibnamefont {Pappa}},\ }\href@noop {} {\bibfield  {journal} {\bibinfo  {journal} {PRX Quantum}\ }\textbf {\bibinfo {volume} {1}},\ \bibinfo {pages} {020325} (\bibinfo {year} {2020})}\BibitemShut {NoStop}%
\bibitem [{\citenamefont {Thalacker}\ \emph {et~al.}(2021)\citenamefont {Thalacker}, \citenamefont {Hahn}, \citenamefont {de~Jong}, \citenamefont {Pappa},\ and\ \citenamefont {Barz}}]{thalacker2021anonymous}%
  \BibitemOpen
  \bibfield  {author} {\bibinfo {author} {\bibfnamefont {C.}~\bibnamefont {Thalacker}}, \bibinfo {author} {\bibfnamefont {F.}~\bibnamefont {Hahn}}, \bibinfo {author} {\bibfnamefont {J.}~\bibnamefont {de~Jong}}, \bibinfo {author} {\bibfnamefont {A.}~\bibnamefont {Pappa}}, \ and\ \bibinfo {author} {\bibfnamefont {S.}~\bibnamefont {Barz}},\ }\href@noop {} {\bibfield  {journal} {\bibinfo  {journal} {New Journal of Physics}\ }\textbf {\bibinfo {volume} {23}},\ \bibinfo {pages} {083026} (\bibinfo {year} {2021})}\BibitemShut {NoStop}%
\bibitem [{\citenamefont {Huang}\ \emph {et~al.}(2022)\citenamefont {Huang}, \citenamefont {Joshi}, \citenamefont {Aktas}, \citenamefont {Lupo}, \citenamefont {Quintavalle}, \citenamefont {Venkatachalam}, \citenamefont {Wengerowsky}, \citenamefont {Lon{\v{c}}ari{\'c}}, \citenamefont {Neumann}, \citenamefont {Liu} \emph {et~al.}}]{huang2022experimental}%
  \BibitemOpen
  \bibfield  {author} {\bibinfo {author} {\bibfnamefont {Z.}~\bibnamefont {Huang}}, \bibinfo {author} {\bibfnamefont {S.~K.}\ \bibnamefont {Joshi}}, \bibinfo {author} {\bibfnamefont {D.}~\bibnamefont {Aktas}}, \bibinfo {author} {\bibfnamefont {C.}~\bibnamefont {Lupo}}, \bibinfo {author} {\bibfnamefont {A.~O.}\ \bibnamefont {Quintavalle}}, \bibinfo {author} {\bibfnamefont {N.}~\bibnamefont {Venkatachalam}}, \bibinfo {author} {\bibfnamefont {S.}~\bibnamefont {Wengerowsky}}, \bibinfo {author} {\bibfnamefont {M.}~\bibnamefont {Lon{\v{c}}ari{\'c}}}, \bibinfo {author} {\bibfnamefont {S.~P.}\ \bibnamefont {Neumann}}, \bibinfo {author} {\bibfnamefont {B.}~\bibnamefont {Liu}},  \emph {et~al.},\ }\href@noop {} {\bibfield  {journal} {\bibinfo  {journal} {npj Quantum Information}\ }\textbf {\bibinfo {volume} {8}},\ \bibinfo {pages} {25} (\bibinfo {year} {2022})}\BibitemShut {NoStop}%
\bibitem [{\citenamefont {Grasselli}\ \emph {et~al.}(2022)\citenamefont {Grasselli}, \citenamefont {Murta}, \citenamefont {de~Jong}, \citenamefont {Hahn}, \citenamefont {Bru{\ss}}, \citenamefont {Kampermann},\ and\ \citenamefont {Pappa}}]{Grasselli2022}%
  \BibitemOpen
  \bibfield  {author} {\bibinfo {author} {\bibfnamefont {F.}~\bibnamefont {Grasselli}}, \bibinfo {author} {\bibfnamefont {G.}~\bibnamefont {Murta}}, \bibinfo {author} {\bibfnamefont {J.}~\bibnamefont {de~Jong}}, \bibinfo {author} {\bibfnamefont {F.}~\bibnamefont {Hahn}}, \bibinfo {author} {\bibfnamefont {D.}~\bibnamefont {Bru{\ss}}}, \bibinfo {author} {\bibfnamefont {H.}~\bibnamefont {Kampermann}}, \ and\ \bibinfo {author} {\bibfnamefont {A.}~\bibnamefont {Pappa}},\ }\href@noop {} {\bibfield  {journal} {\bibinfo  {journal} {PRX Quantum}\ }\textbf {\bibinfo {volume} {3}},\ \bibinfo {pages} {040306} (\bibinfo {year} {2022})}\BibitemShut {NoStop}%
\bibitem [{\citenamefont {de~Jong}\ \emph {et~al.}(2023)\citenamefont {de~Jong}, \citenamefont {Hahn}, \citenamefont {Eisert}, \citenamefont {Walk},\ and\ \citenamefont {Pappa}}]{de2023anonymous}%
  \BibitemOpen
  \bibfield  {author} {\bibinfo {author} {\bibfnamefont {J.}~\bibnamefont {de~Jong}}, \bibinfo {author} {\bibfnamefont {F.}~\bibnamefont {Hahn}}, \bibinfo {author} {\bibfnamefont {J.}~\bibnamefont {Eisert}}, \bibinfo {author} {\bibfnamefont {N.}~\bibnamefont {Walk}}, \ and\ \bibinfo {author} {\bibfnamefont {A.}~\bibnamefont {Pappa}},\ }\href@noop {} {\bibfield  {journal} {\bibinfo  {journal} {Quantum}\ }\textbf {\bibinfo {volume} {7}},\ \bibinfo {pages} {1117} (\bibinfo {year} {2023})}\BibitemShut {NoStop}%
\bibitem [{\citenamefont {Webb}\ \emph {et~al.}(2024)\citenamefont {Webb}, \citenamefont {Ho}, \citenamefont {Grasselli}, \citenamefont {Murta}, \citenamefont {Pickston}, \citenamefont {Ulibarrena},\ and\ \citenamefont {Fedrizzi}}]{Webb24}%
  \BibitemOpen
  \bibfield  {author} {\bibinfo {author} {\bibfnamefont {J.~W.}\ \bibnamefont {Webb}}, \bibinfo {author} {\bibfnamefont {J.}~\bibnamefont {Ho}}, \bibinfo {author} {\bibfnamefont {F.}~\bibnamefont {Grasselli}}, \bibinfo {author} {\bibfnamefont {G.}~\bibnamefont {Murta}}, \bibinfo {author} {\bibfnamefont {A.}~\bibnamefont {Pickston}}, \bibinfo {author} {\bibfnamefont {A.}~\bibnamefont {Ulibarrena}}, \ and\ \bibinfo {author} {\bibfnamefont {A.}~\bibnamefont {Fedrizzi}},\ }\href@noop {} {\bibfield  {journal} {\bibinfo  {journal} {Optica}\ }\textbf {\bibinfo {volume} {11}},\ \bibinfo {pages} {872} (\bibinfo {year} {2024})}\BibitemShut {NoStop}%
\bibitem [{\citenamefont {Eldredge}\ \emph {et~al.}(2018)\citenamefont {Eldredge}, \citenamefont {Foss-Feig}, \citenamefont {Gross}, \citenamefont {Rolston},\ and\ \citenamefont {Gorshkov}}]{eldredge2018optimal}%
  \BibitemOpen
  \bibfield  {author} {\bibinfo {author} {\bibfnamefont {Z.}~\bibnamefont {Eldredge}}, \bibinfo {author} {\bibfnamefont {M.}~\bibnamefont {Foss-Feig}}, \bibinfo {author} {\bibfnamefont {J.~A.}\ \bibnamefont {Gross}}, \bibinfo {author} {\bibfnamefont {S.~L.}\ \bibnamefont {Rolston}}, \ and\ \bibinfo {author} {\bibfnamefont {A.~V.}\ \bibnamefont {Gorshkov}},\ }\href@noop {} {\bibfield  {journal} {\bibinfo  {journal} {Physical Review A}\ }\textbf {\bibinfo {volume} {97}},\ \bibinfo {pages} {042337} (\bibinfo {year} {2018})}\BibitemShut {NoStop}%
\bibitem [{\citenamefont {Huang}\ \emph {et~al.}(2019)\citenamefont {Huang}, \citenamefont {Macchiavello},\ and\ \citenamefont {Maccone}}]{Huang2019}%
  \BibitemOpen
  \bibfield  {author} {\bibinfo {author} {\bibfnamefont {Z.}~\bibnamefont {Huang}}, \bibinfo {author} {\bibfnamefont {C.}~\bibnamefont {Macchiavello}}, \ and\ \bibinfo {author} {\bibfnamefont {L.}~\bibnamefont {Maccone}},\ }\href@noop {} {\bibfield  {journal} {\bibinfo  {journal} {Physical Review A}\ }\textbf {\bibinfo {volume} {99}},\ \bibinfo {pages} {022314} (\bibinfo {year} {2019})}\BibitemShut {NoStop}%
\bibitem [{\citenamefont {Shettell}\ \emph {et~al.}(2022{\natexlab{a}})\citenamefont {Shettell}, \citenamefont {Kashefi},\ and\ \citenamefont {Markham}}]{Shettell2022cryptographicMetrology}%
  \BibitemOpen
  \bibfield  {author} {\bibinfo {author} {\bibfnamefont {N.}~\bibnamefont {Shettell}}, \bibinfo {author} {\bibfnamefont {E.}~\bibnamefont {Kashefi}}, \ and\ \bibinfo {author} {\bibfnamefont {D.}~\bibnamefont {Markham}},\ }\href@noop {} {\bibfield  {journal} {\bibinfo  {journal} {Physical Review A}\ }\textbf {\bibinfo {volume} {105}},\ \bibinfo {pages} {L010401} (\bibinfo {year} {2022}{\natexlab{a}})}\BibitemShut {NoStop}%
\bibitem [{\citenamefont {Kasai}\ \emph {et~al.}(2023)\citenamefont {Kasai}, \citenamefont {Takeuchi}, \citenamefont {Matsuzaki},\ and\ \citenamefont {Tokura}}]{Kasai2023}%
  \BibitemOpen
  \bibfield  {author} {\bibinfo {author} {\bibfnamefont {H.}~\bibnamefont {Kasai}}, \bibinfo {author} {\bibfnamefont {Y.}~\bibnamefont {Takeuchi}}, \bibinfo {author} {\bibfnamefont {Y.}~\bibnamefont {Matsuzaki}}, \ and\ \bibinfo {author} {\bibfnamefont {Y.}~\bibnamefont {Tokura}},\ }\href@noop {} {\bibfield  {journal} {\bibinfo  {journal} {arXiv preprint: 2305.14119}\ } (\bibinfo {year} {2023})}\BibitemShut {NoStop}%
\bibitem [{\citenamefont {Zhang}\ and\ \citenamefont {Zhuang}(2021)}]{zhang2021distributedRev}%
  \BibitemOpen
  \bibfield  {author} {\bibinfo {author} {\bibfnamefont {Z.}~\bibnamefont {Zhang}}\ and\ \bibinfo {author} {\bibfnamefont {Q.}~\bibnamefont {Zhuang}},\ }\href@noop {} {\bibfield  {journal} {\bibinfo  {journal} {Quantum Science and Technology}\ }\textbf {\bibinfo {volume} {6}},\ \bibinfo {pages} {043001} (\bibinfo {year} {2021})}\BibitemShut {NoStop}%
\bibitem [{\citenamefont {Knott}\ \emph {et~al.}(2016)\citenamefont {Knott}, \citenamefont {Proctor}, \citenamefont {Hayes}, \citenamefont {Ralph}, \citenamefont {Kok},\ and\ \citenamefont {Dunningham}}]{knott2016multiparameter}%
  \BibitemOpen
  \bibfield  {author} {\bibinfo {author} {\bibfnamefont {P.~A.}\ \bibnamefont {Knott}}, \bibinfo {author} {\bibfnamefont {T.~J.}\ \bibnamefont {Proctor}}, \bibinfo {author} {\bibfnamefont {A.~J.}\ \bibnamefont {Hayes}}, \bibinfo {author} {\bibfnamefont {J.~F.}\ \bibnamefont {Ralph}}, \bibinfo {author} {\bibfnamefont {P.}~\bibnamefont {Kok}}, \ and\ \bibinfo {author} {\bibfnamefont {J.~A.}\ \bibnamefont {Dunningham}},\ }\href@noop {} {\bibfield  {journal} {\bibinfo  {journal} {Physical Review A}\ }\textbf {\bibinfo {volume} {94}},\ \bibinfo {pages} {062312} (\bibinfo {year} {2016})}\BibitemShut {NoStop}%
\bibitem [{\citenamefont {Mamin}\ \emph {et~al.}(2013)\citenamefont {Mamin}, \citenamefont {Kim}, \citenamefont {Sherwood}, \citenamefont {Rettner}, \citenamefont {Ohno}, \citenamefont {Awschalom},\ and\ \citenamefont {Rugar}}]{mamin2013nanoscale}%
  \BibitemOpen
  \bibfield  {author} {\bibinfo {author} {\bibfnamefont {H.}~\bibnamefont {Mamin}}, \bibinfo {author} {\bibfnamefont {M.}~\bibnamefont {Kim}}, \bibinfo {author} {\bibfnamefont {M.}~\bibnamefont {Sherwood}}, \bibinfo {author} {\bibfnamefont {C.~T.}\ \bibnamefont {Rettner}}, \bibinfo {author} {\bibfnamefont {K.}~\bibnamefont {Ohno}}, \bibinfo {author} {\bibfnamefont {D.}~\bibnamefont {Awschalom}}, \ and\ \bibinfo {author} {\bibfnamefont {D.}~\bibnamefont {Rugar}},\ }\href@noop {} {\bibfield  {journal} {\bibinfo  {journal} {Science}\ }\textbf {\bibinfo {volume} {339}},\ \bibinfo {pages} {557} (\bibinfo {year} {2013})}\BibitemShut {NoStop}%
\bibitem [{\citenamefont {Xia}\ \emph {et~al.}(2020)\citenamefont {Xia}, \citenamefont {Li}, \citenamefont {Clark}, \citenamefont {Hart}, \citenamefont {Zhuang},\ and\ \citenamefont {Zhang}}]{xia2020demonstration}%
  \BibitemOpen
  \bibfield  {author} {\bibinfo {author} {\bibfnamefont {Y.}~\bibnamefont {Xia}}, \bibinfo {author} {\bibfnamefont {W.}~\bibnamefont {Li}}, \bibinfo {author} {\bibfnamefont {W.}~\bibnamefont {Clark}}, \bibinfo {author} {\bibfnamefont {D.}~\bibnamefont {Hart}}, \bibinfo {author} {\bibfnamefont {Q.}~\bibnamefont {Zhuang}}, \ and\ \bibinfo {author} {\bibfnamefont {Z.}~\bibnamefont {Zhang}},\ }\href@noop {} {\bibfield  {journal} {\bibinfo  {journal} {Physical Review Letters}\ }\textbf {\bibinfo {volume} {124}},\ \bibinfo {pages} {150502} (\bibinfo {year} {2020})}\BibitemShut {NoStop}%
\bibitem [{\citenamefont {Ge}\ \emph {et~al.}(2018)\citenamefont {Ge}, \citenamefont {Jacobs}, \citenamefont {Eldredge}, \citenamefont {Gorshkov},\ and\ \citenamefont {Foss-Feig}}]{ge2018distributed}%
  \BibitemOpen
  \bibfield  {author} {\bibinfo {author} {\bibfnamefont {W.}~\bibnamefont {Ge}}, \bibinfo {author} {\bibfnamefont {K.}~\bibnamefont {Jacobs}}, \bibinfo {author} {\bibfnamefont {Z.}~\bibnamefont {Eldredge}}, \bibinfo {author} {\bibfnamefont {A.~V.}\ \bibnamefont {Gorshkov}}, \ and\ \bibinfo {author} {\bibfnamefont {M.}~\bibnamefont {Foss-Feig}},\ }\href@noop {} {\bibfield  {journal} {\bibinfo  {journal} {Physical review letters}\ }\textbf {\bibinfo {volume} {121}},\ \bibinfo {pages} {043604} (\bibinfo {year} {2018})}\BibitemShut {NoStop}%
\bibitem [{\citenamefont {Zhuang}\ \emph {et~al.}(2018)\citenamefont {Zhuang}, \citenamefont {Zhang},\ and\ \citenamefont {Shapiro}}]{zhuang2018distributed}%
  \BibitemOpen
  \bibfield  {author} {\bibinfo {author} {\bibfnamefont {Q.}~\bibnamefont {Zhuang}}, \bibinfo {author} {\bibfnamefont {Z.}~\bibnamefont {Zhang}}, \ and\ \bibinfo {author} {\bibfnamefont {J.~H.}\ \bibnamefont {Shapiro}},\ }\href@noop {} {\bibfield  {journal} {\bibinfo  {journal} {Physical Review A}\ }\textbf {\bibinfo {volume} {97}},\ \bibinfo {pages} {032329} (\bibinfo {year} {2018})}\BibitemShut {NoStop}%
\bibitem [{\citenamefont {Proctor}\ \emph {et~al.}(2018)\citenamefont {Proctor}, \citenamefont {Knott},\ and\ \citenamefont {Dunningham}}]{proctor2018multiparameter}%
  \BibitemOpen
  \bibfield  {author} {\bibinfo {author} {\bibfnamefont {T.~J.}\ \bibnamefont {Proctor}}, \bibinfo {author} {\bibfnamefont {P.~A.}\ \bibnamefont {Knott}}, \ and\ \bibinfo {author} {\bibfnamefont {J.~A.}\ \bibnamefont {Dunningham}},\ }\href@noop {} {\bibfield  {journal} {\bibinfo  {journal} {Physical Review Letters}\ }\textbf {\bibinfo {volume} {120}},\ \bibinfo {pages} {080501} (\bibinfo {year} {2018})}\BibitemShut {NoStop}%
\bibitem [{\citenamefont {Qian}\ \emph {et~al.}(2019)\citenamefont {Qian}, \citenamefont {Eldredge}, \citenamefont {Ge}, \citenamefont {Pagano}, \citenamefont {Monroe}, \citenamefont {Porto},\ and\ \citenamefont {Gorshkov}}]{qian2019heisenberg}%
  \BibitemOpen
  \bibfield  {author} {\bibinfo {author} {\bibfnamefont {K.}~\bibnamefont {Qian}}, \bibinfo {author} {\bibfnamefont {Z.}~\bibnamefont {Eldredge}}, \bibinfo {author} {\bibfnamefont {W.}~\bibnamefont {Ge}}, \bibinfo {author} {\bibfnamefont {G.}~\bibnamefont {Pagano}}, \bibinfo {author} {\bibfnamefont {C.}~\bibnamefont {Monroe}}, \bibinfo {author} {\bibfnamefont {J.~V.}\ \bibnamefont {Porto}}, \ and\ \bibinfo {author} {\bibfnamefont {A.~V.}\ \bibnamefont {Gorshkov}},\ }\href@noop {} {\bibfield  {journal} {\bibinfo  {journal} {Physical Review A}\ }\textbf {\bibinfo {volume} {100}},\ \bibinfo {pages} {042304} (\bibinfo {year} {2019})}\BibitemShut {NoStop}%
\bibitem [{\citenamefont {Takeuchi}\ \emph {et~al.}(2019)\citenamefont {Takeuchi}, \citenamefont {Matsuzaki}, \citenamefont {Miyanishi}, \citenamefont {Sugiyama},\ and\ \citenamefont {Munro}}]{Takeuchi2019}%
  \BibitemOpen
  \bibfield  {author} {\bibinfo {author} {\bibfnamefont {Y.}~\bibnamefont {Takeuchi}}, \bibinfo {author} {\bibfnamefont {Y.}~\bibnamefont {Matsuzaki}}, \bibinfo {author} {\bibfnamefont {K.}~\bibnamefont {Miyanishi}}, \bibinfo {author} {\bibfnamefont {T.}~\bibnamefont {Sugiyama}}, \ and\ \bibinfo {author} {\bibfnamefont {W.~J.}\ \bibnamefont {Munro}},\ }\href@noop {} {\bibfield  {journal} {\bibinfo  {journal} {Physical Review A}\ }\textbf {\bibinfo {volume} {99}},\ \bibinfo {pages} {022325} (\bibinfo {year} {2019})}\BibitemShut {NoStop}%
\bibitem [{\citenamefont {Yin}\ \emph {et~al.}(2020)\citenamefont {Yin}, \citenamefont {Takeuchi}, \citenamefont {Zhang}, \citenamefont {Yin}, \citenamefont {Matsuzaki}, \citenamefont {Peng}, \citenamefont {Xu}, \citenamefont {Xu}, \citenamefont {Tang}, \citenamefont {Zhou}, \citenamefont {Chen}, \citenamefont {Li},\ and\ \citenamefont {Guo}}]{Yin2020}%
  \BibitemOpen
  \bibfield  {author} {\bibinfo {author} {\bibfnamefont {P.}~\bibnamefont {Yin}}, \bibinfo {author} {\bibfnamefont {Y.}~\bibnamefont {Takeuchi}}, \bibinfo {author} {\bibfnamefont {W.-H.}\ \bibnamefont {Zhang}}, \bibinfo {author} {\bibfnamefont {Z.-Q.}\ \bibnamefont {Yin}}, \bibinfo {author} {\bibfnamefont {Y.}~\bibnamefont {Matsuzaki}}, \bibinfo {author} {\bibfnamefont {X.-X.}\ \bibnamefont {Peng}}, \bibinfo {author} {\bibfnamefont {X.-Y.}\ \bibnamefont {Xu}}, \bibinfo {author} {\bibfnamefont {J.-S.}\ \bibnamefont {Xu}}, \bibinfo {author} {\bibfnamefont {J.-S.}\ \bibnamefont {Tang}}, \bibinfo {author} {\bibfnamefont {Z.-Q.}\ \bibnamefont {Zhou}}, \bibinfo {author} {\bibfnamefont {G.}~\bibnamefont {Chen}}, \bibinfo {author} {\bibfnamefont {C.-F.}\ \bibnamefont {Li}}, \ and\ \bibinfo {author} {\bibfnamefont {G.-C.}\ \bibnamefont {Guo}},\ }\href@noop {} {\bibfield  {journal} {\bibinfo  {journal} {Physical Review Applied}\ }\textbf {\bibinfo {volume} {14}},\ \bibinfo {pages} {014065} (\bibinfo {year}
  {2020})}\BibitemShut {NoStop}%
\bibitem [{\citenamefont {Shettell}\ \emph {et~al.}(2022{\natexlab{b}})\citenamefont {Shettell}, \citenamefont {Hassani},\ and\ \citenamefont {Markham}}]{Shettell2022arxiv}%
  \BibitemOpen
  \bibfield  {author} {\bibinfo {author} {\bibfnamefont {N.}~\bibnamefont {Shettell}}, \bibinfo {author} {\bibfnamefont {M.}~\bibnamefont {Hassani}}, \ and\ \bibinfo {author} {\bibfnamefont {D.}~\bibnamefont {Markham}},\ }\href@noop {} {\bibfield  {journal} {\bibinfo  {journal} {arXiv preprint: 2207.14450}\ } (\bibinfo {year} {2022}{\natexlab{b}})}\BibitemShut {NoStop}%
\bibitem [{\citenamefont {Proietti}\ \emph {et~al.}(2021)\citenamefont {Proietti}, \citenamefont {Ho}, \citenamefont {Grasselli}, \citenamefont {Barrow}, \citenamefont {Malik},\ and\ \citenamefont {Fedrizzi}}]{Proietti2021}%
  \BibitemOpen
  \bibfield  {author} {\bibinfo {author} {\bibfnamefont {M.}~\bibnamefont {Proietti}}, \bibinfo {author} {\bibfnamefont {J.}~\bibnamefont {Ho}}, \bibinfo {author} {\bibfnamefont {F.}~\bibnamefont {Grasselli}}, \bibinfo {author} {\bibfnamefont {P.}~\bibnamefont {Barrow}}, \bibinfo {author} {\bibfnamefont {M.}~\bibnamefont {Malik}}, \ and\ \bibinfo {author} {\bibfnamefont {A.}~\bibnamefont {Fedrizzi}},\ }\href@noop {} {\bibfield  {journal} {\bibinfo  {journal} {Science Advances}\ }\textbf {\bibinfo {volume} {7}},\ \bibinfo {pages} {eabe0395} (\bibinfo {year} {2021})}\BibitemShut {NoStop}%
\bibitem [{\citenamefont {Pickston}\ \emph {et~al.}(2023)\citenamefont {Pickston}, \citenamefont {Ho}, \citenamefont {Ulibarrena}, \citenamefont {Grasselli}, \citenamefont {Proietti}, \citenamefont {Morrison}, \citenamefont {Barrow}, \citenamefont {Graffitti},\ and\ \citenamefont {Fedrizzi}}]{pickston2023conference}%
  \BibitemOpen
  \bibfield  {author} {\bibinfo {author} {\bibfnamefont {A.}~\bibnamefont {Pickston}}, \bibinfo {author} {\bibfnamefont {J.}~\bibnamefont {Ho}}, \bibinfo {author} {\bibfnamefont {A.}~\bibnamefont {Ulibarrena}}, \bibinfo {author} {\bibfnamefont {F.}~\bibnamefont {Grasselli}}, \bibinfo {author} {\bibfnamefont {M.}~\bibnamefont {Proietti}}, \bibinfo {author} {\bibfnamefont {C.~L.}\ \bibnamefont {Morrison}}, \bibinfo {author} {\bibfnamefont {P.}~\bibnamefont {Barrow}}, \bibinfo {author} {\bibfnamefont {F.}~\bibnamefont {Graffitti}}, \ and\ \bibinfo {author} {\bibfnamefont {A.}~\bibnamefont {Fedrizzi}},\ }\href@noop {} {\bibfield  {journal} {\bibinfo  {journal} {npj Quantum Information}\ }\textbf {\bibinfo {volume} {9}},\ \bibinfo {pages} {82} (\bibinfo {year} {2023})}\BibitemShut {NoStop}%
\bibitem [{\citenamefont {Murta}\ \emph {et~al.}(2020)\citenamefont {Murta}, \citenamefont {Grasselli}, \citenamefont {Kampermann},\ and\ \citenamefont {Bru{\ss}}}]{Murta2020}%
  \BibitemOpen
  \bibfield  {author} {\bibinfo {author} {\bibfnamefont {G.}~\bibnamefont {Murta}}, \bibinfo {author} {\bibfnamefont {F.}~\bibnamefont {Grasselli}}, \bibinfo {author} {\bibfnamefont {H.}~\bibnamefont {Kampermann}}, \ and\ \bibinfo {author} {\bibfnamefont {D.}~\bibnamefont {Bru{\ss}}},\ }\href@noop {} {\bibfield  {journal} {\bibinfo  {journal} {Advanced Quantum Technologies}\ }\textbf {\bibinfo {volume} {3}},\ \bibinfo {pages} {2000025} (\bibinfo {year} {2020})}\BibitemShut {NoStop}%
\bibitem [{\citenamefont {T{\'o}th}\ and\ \citenamefont {G{\"u}hne}(2005)}]{toth2005entanglement}%
  \BibitemOpen
  \bibfield  {author} {\bibinfo {author} {\bibfnamefont {G.}~\bibnamefont {T{\'o}th}}\ and\ \bibinfo {author} {\bibfnamefont {O.}~\bibnamefont {G{\"u}hne}},\ }\href@noop {} {\bibfield  {journal} {\bibinfo  {journal} {Physical Review A}\ }\textbf {\bibinfo {volume} {72}},\ \bibinfo {pages} {022340} (\bibinfo {year} {2005})}\BibitemShut {NoStop}%
\bibitem [{\citenamefont {Unnikrishnan}\ and\ \citenamefont {Markham}(2022)}]{unnikrishnan2022verification}%
  \BibitemOpen
  \bibfield  {author} {\bibinfo {author} {\bibfnamefont {A.}~\bibnamefont {Unnikrishnan}}\ and\ \bibinfo {author} {\bibfnamefont {D.}~\bibnamefont {Markham}},\ }\href@noop {} {\bibfield  {journal} {\bibinfo  {journal} {Physical Review A}\ }\textbf {\bibinfo {volume} {105}},\ \bibinfo {pages} {052420} (\bibinfo {year} {2022})}\BibitemShut {NoStop}%
\bibitem [{\citenamefont {Rezakhani}\ \emph {et~al.}(2019{\natexlab{a}})\citenamefont {Rezakhani}, \citenamefont {Hassani},\ and\ \citenamefont {Alipour}}]{continuityFisherInfo}%
  \BibitemOpen
  \bibfield  {author} {\bibinfo {author} {\bibfnamefont {A.~T.}\ \bibnamefont {Rezakhani}}, \bibinfo {author} {\bibfnamefont {M.}~\bibnamefont {Hassani}}, \ and\ \bibinfo {author} {\bibfnamefont {S.}~\bibnamefont {Alipour}},\ }\href {\doibase 10.1103/PhysRevA.100.032317} {\bibfield  {journal} {\bibinfo  {journal} {Phys. Rev. A}\ }\textbf {\bibinfo {volume} {100}},\ \bibinfo {pages} {032317} (\bibinfo {year} {2019}{\natexlab{a}})}\BibitemShut {NoStop}%
\bibitem [{\citenamefont {Pickston}\ \emph {et~al.}(2021)\citenamefont {Pickston}, \citenamefont {Graffitti}, \citenamefont {Barrow}, \citenamefont {Morrison}, \citenamefont {Ho}, \citenamefont {Bra{\'n}czyk},\ and\ \citenamefont {Fedrizzi}}]{pickston2021optimised}%
  \BibitemOpen
  \bibfield  {author} {\bibinfo {author} {\bibfnamefont {A.}~\bibnamefont {Pickston}}, \bibinfo {author} {\bibfnamefont {F.}~\bibnamefont {Graffitti}}, \bibinfo {author} {\bibfnamefont {P.}~\bibnamefont {Barrow}}, \bibinfo {author} {\bibfnamefont {C.~L.}\ \bibnamefont {Morrison}}, \bibinfo {author} {\bibfnamefont {J.}~\bibnamefont {Ho}}, \bibinfo {author} {\bibfnamefont {A.~M.}\ \bibnamefont {Bra{\'n}czyk}}, \ and\ \bibinfo {author} {\bibfnamefont {A.}~\bibnamefont {Fedrizzi}},\ }\href@noop {} {\bibfield  {journal} {\bibinfo  {journal} {Optics Express}\ }\textbf {\bibinfo {volume} {29}},\ \bibinfo {pages} {6991} (\bibinfo {year} {2021})}\BibitemShut {NoStop}%
\bibitem [{\citenamefont {Fedrizzi}\ \emph {et~al.}(2007)\citenamefont {Fedrizzi}, \citenamefont {Herbst}, \citenamefont {Poppe}, \citenamefont {Jennewein},\ and\ \citenamefont {Zeilinger}}]{fedrizzi2007}%
  \BibitemOpen
  \bibfield  {author} {\bibinfo {author} {\bibfnamefont {A.}~\bibnamefont {Fedrizzi}}, \bibinfo {author} {\bibfnamefont {T.}~\bibnamefont {Herbst}}, \bibinfo {author} {\bibfnamefont {A.}~\bibnamefont {Poppe}}, \bibinfo {author} {\bibfnamefont {T.}~\bibnamefont {Jennewein}}, \ and\ \bibinfo {author} {\bibfnamefont {A.}~\bibnamefont {Zeilinger}},\ }\href@noop {} {\bibfield  {journal} {\bibinfo  {journal} {Optics Express}\ }\textbf {\bibinfo {volume} {15}},\ \bibinfo {pages} {15377} (\bibinfo {year} {2007})}\BibitemShut {NoStop}%
\bibitem [{\citenamefont {Colisson}\ \emph {et~al.}(2024)\citenamefont {Colisson}, \citenamefont {Markham},\ and\ \citenamefont {Yehia}}]{Colisson2024}%
  \BibitemOpen
  \bibfield  {author} {\bibinfo {author} {\bibfnamefont {L.}~\bibnamefont {Colisson}}, \bibinfo {author} {\bibfnamefont {D.}~\bibnamefont {Markham}}, \ and\ \bibinfo {author} {\bibfnamefont {R.}~\bibnamefont {Yehia}},\ }\href@noop {} {\bibfield  {journal} {\bibinfo  {journal} {arXiv preprint: 2402.01445}\ } (\bibinfo {year} {2024})}\BibitemShut {NoStop}%
\bibitem [{\citenamefont {Unnikrishnan}\ \emph {et~al.}(2019)\citenamefont {Unnikrishnan}, \citenamefont {MacFarlane}, \citenamefont {Yi}, \citenamefont {Diamanti}, \citenamefont {Markham},\ and\ \citenamefont {Kerenidis}}]{Unnikrishnan2019}%
  \BibitemOpen
  \bibfield  {author} {\bibinfo {author} {\bibfnamefont {A.}~\bibnamefont {Unnikrishnan}}, \bibinfo {author} {\bibfnamefont {I.~J.}\ \bibnamefont {MacFarlane}}, \bibinfo {author} {\bibfnamefont {R.}~\bibnamefont {Yi}}, \bibinfo {author} {\bibfnamefont {E.}~\bibnamefont {Diamanti}}, \bibinfo {author} {\bibfnamefont {D.}~\bibnamefont {Markham}}, \ and\ \bibinfo {author} {\bibfnamefont {I.}~\bibnamefont {Kerenidis}},\ }\href@noop {} {\bibfield  {journal} {\bibinfo  {journal} {Physical Review Letters}\ }\textbf {\bibinfo {volume} {122}},\ \bibinfo {pages} {240501} (\bibinfo {year} {2019})}\BibitemShut {NoStop}%
\bibitem [{\citenamefont {Go{\v{c}}anin}\ \emph {et~al.}(2022)\citenamefont {Go{\v{c}}anin}, \citenamefont {{\v{S}}upi{\'c}},\ and\ \citenamefont {Daki{\'c}}}]{govcanin2022sample}%
  \BibitemOpen
  \bibfield  {author} {\bibinfo {author} {\bibfnamefont {A.}~\bibnamefont {Go{\v{c}}anin}}, \bibinfo {author} {\bibfnamefont {I.}~\bibnamefont {{\v{S}}upi{\'c}}}, \ and\ \bibinfo {author} {\bibfnamefont {B.}~\bibnamefont {Daki{\'c}}},\ }\href@noop {} {\bibfield  {journal} {\bibinfo  {journal} {PRX Quantum}\ }\textbf {\bibinfo {volume} {3}},\ \bibinfo {pages} {010317} (\bibinfo {year} {2022})}\BibitemShut {NoStop}%
\bibitem [{\citenamefont {Antesberger}\ \emph {et~al.}(2024)\citenamefont {Antesberger}, \citenamefont {Schmid}, \citenamefont {Cao}, \citenamefont {Daki{\'c}}, \citenamefont {Rozema},\ and\ \citenamefont {Walther}}]{antesberger2024efficient}%
  \BibitemOpen
  \bibfield  {author} {\bibinfo {author} {\bibfnamefont {M.}~\bibnamefont {Antesberger}}, \bibinfo {author} {\bibfnamefont {M.~M.}\ \bibnamefont {Schmid}}, \bibinfo {author} {\bibfnamefont {H.}~\bibnamefont {Cao}}, \bibinfo {author} {\bibfnamefont {B.}~\bibnamefont {Daki{\'c}}}, \bibinfo {author} {\bibfnamefont {L.~A.}\ \bibnamefont {Rozema}}, \ and\ \bibinfo {author} {\bibfnamefont {P.}~\bibnamefont {Walther}},\ }\href@noop {} {\bibfield  {journal} {\bibinfo  {journal} {arXiv preprint:2407.13913}\ } (\bibinfo {year} {2024})}\BibitemShut {NoStop}%
\bibitem [{\citenamefont {Martins}\ \emph {et~al.}(2024)\citenamefont {Martins}, \citenamefont {Laurent-Puig}, \citenamefont {{\v{S}}upi{\'c}}, \citenamefont {Markham},\ and\ \citenamefont {Diamanti}}]{martins2024experimental}%
  \BibitemOpen
  \bibfield  {author} {\bibinfo {author} {\bibfnamefont {L.~d.~S.}\ \bibnamefont {Martins}}, \bibinfo {author} {\bibfnamefont {N.}~\bibnamefont {Laurent-Puig}}, \bibinfo {author} {\bibfnamefont {I.}~\bibnamefont {{\v{S}}upi{\'c}}}, \bibinfo {author} {\bibfnamefont {D.}~\bibnamefont {Markham}}, \ and\ \bibinfo {author} {\bibfnamefont {E.}~\bibnamefont {Diamanti}},\ }\href@noop {} {\bibfield  {journal} {\bibinfo  {journal} {arXiv preprint:2407.13529}\ } (\bibinfo {year} {2024})}\BibitemShut {NoStop}%
\bibitem [{\citenamefont {Bugalho}\ \emph {et~al.}(2025)\citenamefont {Bugalho}, \citenamefont {Hassani}, \citenamefont {Omar},\ and\ \citenamefont {Markham}}]{bugalho2025}%
  \BibitemOpen
  \bibfield  {author} {\bibinfo {author} {\bibfnamefont {L.}~\bibnamefont {Bugalho}}, \bibinfo {author} {\bibfnamefont {M.}~\bibnamefont {Hassani}}, \bibinfo {author} {\bibfnamefont {Y.}~\bibnamefont {Omar}}, \ and\ \bibinfo {author} {\bibfnamefont {D.}~\bibnamefont {Markham}},\ }\href@noop {} {\bibfield  {journal} {\bibinfo  {journal} {Quantum}\ }\textbf {\bibinfo {volume} {9}},\ \bibinfo {pages} {1596} (\bibinfo {year} {2025})}\BibitemShut {NoStop}%
\bibitem [{\citenamefont {Hassani}\ \emph {et~al.}(2025)\citenamefont {Hassani}, \citenamefont {Scheiner}, \citenamefont {Paris},\ and\ \citenamefont {Markham}}]{hassani2025}%
  \BibitemOpen
  \bibfield  {author} {\bibinfo {author} {\bibfnamefont {M.}~\bibnamefont {Hassani}}, \bibinfo {author} {\bibfnamefont {S.}~\bibnamefont {Scheiner}}, \bibinfo {author} {\bibfnamefont {M.~G.}\ \bibnamefont {Paris}}, \ and\ \bibinfo {author} {\bibfnamefont {D.}~\bibnamefont {Markham}},\ }\href@noop {} {\bibfield  {journal} {\bibinfo  {journal} {Physical Review Letters}\ }\textbf {\bibinfo {volume} {134}},\ \bibinfo {pages} {030802} (\bibinfo {year} {2025})}\BibitemShut {NoStop}%
\bibitem [{\citenamefont {Ko{\l}ody{\'n}ski}(2016)}]{augusiak2016asymptotic}%
  \BibitemOpen
  \bibfield  {author} {\bibinfo {author} {\bibfnamefont {J.}~\bibnamefont {Ko{\l}ody{\'n}ski}},\ }\href@noop {} {\bibfield  {journal} {\bibinfo  {journal} {Physical Review A}\ }\textbf {\bibinfo {volume} {94}},\ \bibinfo {pages} {012339} (\bibinfo {year} {2016})}\BibitemShut {NoStop}%
\bibitem [{\citenamefont {Rezakhani}\ \emph {et~al.}(2019{\natexlab{b}})\citenamefont {Rezakhani}, \citenamefont {Hassani},\ and\ \citenamefont {Alipour}}]{rezakhani2019}%
  \BibitemOpen
  \bibfield  {author} {\bibinfo {author} {\bibfnamefont {A.~T.}\ \bibnamefont {Rezakhani}}, \bibinfo {author} {\bibfnamefont {M.}~\bibnamefont {Hassani}}, \ and\ \bibinfo {author} {\bibfnamefont {S.}~\bibnamefont {Alipour}},\ }\href@noop {} {\bibfield  {journal} {\bibinfo  {journal} {Phys. Rev. A}\ }\textbf {\bibinfo {volume} {100}},\ \bibinfo {pages} {032317} (\bibinfo {year} {2019}{\natexlab{b}})}\BibitemShut {NoStop}%
\bibitem [{\citenamefont {Fiderer}\ \emph {et~al.}(2019)\citenamefont {Fiderer}, \citenamefont {Fra{\"\i}sse},\ and\ \citenamefont {Braun}}]{Fiderer2019}%
  \BibitemOpen
  \bibfield  {author} {\bibinfo {author} {\bibfnamefont {L.~J.}\ \bibnamefont {Fiderer}}, \bibinfo {author} {\bibfnamefont {J.~M.~E.}\ \bibnamefont {Fra{\"\i}sse}}, \ and\ \bibinfo {author} {\bibfnamefont {D.}~\bibnamefont {Braun}},\ }\href@noop {} {\bibfield  {journal} {\bibinfo  {journal} {Physical review letters}\ }\textbf {\bibinfo {volume} {123}},\ \bibinfo {pages} {250502} (\bibinfo {year} {2019})}\BibitemShut {NoStop}%
\bibitem [{\citenamefont {Paris}(2009)}]{ParisQFI}%
  \BibitemOpen
  \bibfield  {author} {\bibinfo {author} {\bibfnamefont {M.~G.~A.}\ \bibnamefont {Paris}},\ }\href@noop {} {\bibfield  {journal} {\bibinfo  {journal} {International Journal of Quantum Information}\ }\textbf {\bibinfo {volume} {07}},\ \bibinfo {pages} {125} (\bibinfo {year} {2009})}\BibitemShut {NoStop}%
\end{thebibliography}
\end{document}